\documentclass{aa}  
\usepackage{graphicx}
\usepackage{txfonts}
\usepackage{natbib}
\def\aaps{A\&AS}
\def\aap{A\&A}
\def\apj{ApJ}
\def\apjs{ApJS}
\def\aj{AJ}

\def\ha{H\,$\alpha$}
\def \hi {\ion{H}{i}}
\def \hii {\ion{H}{ii}~}

\def\kms{km\,s$^{-1}$}
\def\kmss{km\,s$^{-1}~$}
\def\msun{M$_{\sun}$}
\def\msuns{M$_{\sun~}$}
\def\rsun{R$_{\sun}$}
\def\rsuns{R$_{\sun}~$}

\def\vsuns{v$_{\sun}~$}
\def\deg{\hbox{$^\circ$}}

\def\fdg{\hbox{$.\!\!^\circ$}}

\defcitealias{Kalberla2003}{Paper I}

\begin{document}
   \title{Dark matter in the Milky Way}

   \subtitle{II. the \hi~ gas distribution as a tracer of the gravitational
   potential}

   \author{P.\,M.\,W. Kalberla\inst{1}, L. Dedes\inst{1},
     J. Kerp\inst{1} \& U. Haud\inst{2}}

   \institute{ Argelander-Institut f\"ur Astronomie, Universit\"at
     Bonn\thanks{Founded by merging of the Sternwarte,
     Radioastronomisches Institut and Institut f\"ur Astrophysik und
     Extraterrestrische Forschung der Universit\"at Bonn}, Auf dem
     H\"ugel 71, 53121 Bonn, Germany\\
     \email{pkalberla@astro.uni-bonn.de,ldedes@astro.uni-bonn.de,jkerp@astro.uni-bonn.de}
     \and Tartu Observatory, 61602 Toravere, Estonia\\
     \email{urmas@aai.ee} }
      
   \authorrunning{P.\,M.\,W. Kalberla et al. } 

   \titlerunning{Gravity and the Milky Way \hi~ gas distribution}

   \offprints{P.\,M.\,W. Kalberla}

   \date{Received September 7 2006 / Accepted March 20 2007 }

  \abstract 
  {Gas within a galaxy is forced to establish pressure balance against
  gravitational forces. The shape of an unperturbed gaseous disk can be
  used to constrain dark matter models.   }
  {We derive the 3-D \hi~ volume density distribution for the Milky Way
  out to a galactocentric radius of 40 kpc and a height of 20 kpc to
  constrain the Galactic mass distribution. }
  {We used the Leiden/Argentine/Bonn all sky 21-cm line survey. The
  transformation from brightness temperatures to densities depends on
  the rotation curve. We explored several models, 
  reflecting different dark matter distributions. Each of these models
  was set up to solve the combined Poisson-Boltzmann equation in a
  self-consistent way and optimized to reproduce the
  observed flaring. }
  {Besides a massive extended halo of $M \sim 1.8~ 10^{12}$ \msun, we
  find a self-gravitating dark matter disk with $ M=2$ to $3~
  10^{11}$ \msun, including a dark matter ring at $ 13 < R < 18.5 $ kpc
  with $ M = 2.2 $ to $2.8~ 10^{10}$ \msun. The existence of the ring was
  previously postulated from EGRET data and coincides with a giant
  stellar structure that surrounds the Galaxy. The resulting Milky Way
  rotation curve is flat up to $ R \sim 27$ kpc and slowly decreases
  outwards. The \hi~ gas layer is strongly flaring. The HWHM scale
  height is 60 pc at $ R = 4 $ kpc and increases to $\sim 2700$ pc at $R
  = 40$ kpc. Spiral arms cause a noticeable imprint on the
  gravitational field, at least out to $R = 30$ kpc. }
  {Our mass model supports previous proposals that the giant stellar
  ring structure is due to a merging dwarf galaxy. The fact that the
  majority of the dark matter in the Milky Way for $R \la 40$ kpc can
  be successfully modeled by a self-gravitating isothermal disk raises
  the question of whether this massive disk may have been caused by
  similar merger events in the past. The substructure in the Galactic dark
  matter disk suggests a dissipative nature for the dark matter disk. }
  \maketitle
%

\section{Introduction}

The shape of the Galactic \hi~ disk is interesting for its own sake. Size
matters and we would like to know how the Milky Way compares to other
galaxies. When trying to convert the observed brightness temperature
distribution to a volume density distribution one realizes, however, that
this transformation depends on the shape of the rotation curve. 
Determining the rotation curve needs accurate independent
measurements of velocities and distances for a broad distance range, a
daunting task. The first large-scale map of the \hi~ volume density
distribution in the Milky Way \citep{Westerhout1957} was therefore
generated from a mass model \citep{Schmidt1956}. For this model, the
rotation curve at distances $R \ga 10 $ kpc is falling. 

Observations, however, give opposite results. \citet{Georgelin1976} were
the first to notice a rising rotation curve for a relatively large
sample. They used \ha \ radial velocities and distances of stars that
excite Galactic \hii regions and found an increase of $v_{\rm rot}$ for
$R \ga 10$ kpc. Additional systematical variations in $v_{\rm rot}$ for
the northern and southern parts of the Galaxy led to the question of
whether a rotation curve from an axisymmetric model suffers from
systematic errors.

The discrepancy between mass models and observations has persisted until
recently.  Theoretical mass models predict a falling rotation curve
\citep[e.g.][]{Bahcall1980,Caldwell1981,Rohlfs1981,Haud1989,Dehnen98}
while observations have consistently indicated the opposite
\citep[e.g.][]{Blitz1979,Schneider1983,Fich1989,Merrifield1992,Brand1993,Honma1997}.

The situation is further complicated by the fact that the rotation
curve depends on the choice of Galactic constants.  A high
rotational velocity at the position of the Sun leads to a rising
rotation curve, while the opposite holds for a low velocity.  The
interrelation between Galactic constants, dark matter models, and the shape
of the \hi~ distribution has been studied in a series of papers by
\citet{OM98,OM00,OM01}. These authors constrain the Galactic constants
to $R_{\sun} \la 8 $ kpc and $v_{\sun} \la 200$ \kms.

\citet{Rohlfs1988} considered models with a rising rotation curve for $
11 \la R \la 18 $ kpc but found it difficult to explain this by a dark
corona. Such a solution would result in a high rotational velocity of $
400 \la v_{\rm rot} \la 500 $ \kmss outside the Galactic disk, which appears
highly implausible. In addition, such a dark matter distribution would
disable an exponential stellar disk, which conflicts with
observations.

\citet{Binney1997} discuss the question of whether distance errors may
mimic a rising rotation curve. They propose that most of the tracers
that appear to be at $R\ga 11 $ kpc might actually be concentrated
within a ring at $R \sim 14$ kpc. The resulting rotation curve would be
roughly constant or gently falling. Several groups --
\citet{Newberg2002,Ibata2003,Yanny2003,Martin2005,Martinez2005,Bellazzini2006}
-- have subsequently claimed to have detected stellar streams in the
Milky Way, right at the region of interest. A ring-like mass
concentration within this area plays an important role throughout our
paper.

The IAU-recommended Galactic constants $R_{\sun} = 8.5 $ kpc and
$v_{\sun} = 220$ \kmss fit best with a flat rotation curve
\citep{Fich1989}.  Typical Sb galaxies have nearly flat outer rotation
curves \citep{Sofue2001}. It seems to be an acceptable compromise to
assume that the Milky Way is typical in this respect. For about a decade
now there has been little discussion about the Galactic rotation
curve. A flat rotation curve with $ v_{\rm rot} = 220$ \kmss was used
for the most recent determination of the Galactic volume density
distribution by \citet{Levine2006a}.

The need to obtain better constraints for the radial mass distribution
in the Milky Way has been stressed by many authors. Gas can play a
particularly interesting role as a tracer of the gravitational
potential, provided that its velocity dispersion is position-independent
and that there are no external forces. The inner part of a galactic disk
is dominated by stars. This results in strong gravitational forces $k_z$
perpendicular to the disk. The influence of the stars decreases with
increasing galactocentric distance $R$. That allows the gas, which
withstands the forces by means of internal turbulent motions, to reach
larger scale heights. The \hi~ disks in galaxies are typically three times
more extended than their stellar counterparts. This property, as well as
the fact that the outer parts of galaxies are dominated by dark matter,
highlights the importance of the gaseous disk as a tracer of the mass
distribution \citep{Sofue2001,Combes2002}. The Galactic outskirts are
very interesting in this respect.

Gaseous flaring was first used as a diagnostic of halo properties by
\citet{Olling95}, who showed that the halo of NGC 4244 was highly
flattened. \citet{Becquaert1997} obtained a comparable result for NGC
891. Applying a similar analysis to the Milky Way, however, has led to
conflicting results \citep{OM98,OM00,OM01}. \citet{Narayan2005} 
show that neither an oblate nor a prolate isothermal halo can explain
the Milky Way flaring. They advocate a steeper fall-off for the halo
density.

Surprisingly, little is known about the Galactic gas- and mass
distribution at large radial distances. Our knowledge is limited to
distances $ R \la 22$ kpc \citep[][and references therein]{BM}. The
Galactic outskirts are extremely sensitive to the dark matter
distribution.  The main driver of the present paper is therefore to
explore the mass distribution at large radial distances. A second, even
more severe problem has been pointed out by \citet{Dehnen98}. There are
large observational uncertainties in the mass distribution perpendicular
to the disk. Constraints at $z \ga R$ are required.  The gravitational
acceleration $k_z$ in the solar vicinity is well known from stellar
statistics up to distances of $z \la 1.5$ kpc
\citep{KG91,HF2000,Korchagin2003,HF2004}, but little is known at larger
distances. Figure 11 of \citet{HF2004} impressively demonstrates that
observational constraints are missing from star counts. Extra-planar \hi~
gas, though faint, may be a better probe since it extends to large $z$
distances \citep{Lockman91,Kalb98}.

This paper is organized in the following way: Sect. 2 describes our data
analysis, in particular the methods used to derive the \hi~ flaring. In
Sect. 3 we discuss how mass models affect the derived \hi~ volume
density distribution and flaring. Each mass model is checked for
consistency.  The rotation curve associated with the best-fit model is
discussed in Sect. 4. We consider further consistency checks for our
model, using flaring data from the literature and HIPPARCOS surface
densities in Sects. 5\&6. The warp parameters are summarized in Sect. 7. We
discuss deviations by the Milky Way flaring from the axisymmetric model
and spiral structures in Sect. 8. Uncertainties and possible biases due
to model assumptions are considered in Sect. 9. The results are
summarized in Sect. 10.  We discuss the nature of the dark matter ring
and disk in Sect. 11.


\section{Data analysis and methods}

Our study is based on the Leiden/Argentine/Bonn (LAB) \hi~ line survey
\citep{Kalberla2005}. This survey combines the southern sky survey of
the Instituto Argentino de Radioastronom\'ia (IAR) \citep{Bajaja2005}
with an improved version of the Leiden/Dwingeloo Survey (LDS)
\citep{Atlas1997}. Currently, this is the most sensitive Milky Way \hi~
line survey with the most extensive coverage both spatially and
kinematically. The line profiles have been corrected for spurious
side-lobe emission. Low-level emission features can be interpreted safely
as tracers of Galactic structure out to large distances, and we intend to
explore the outskirts of the Milky Way.

\subsection{$T_B(l,b,v)$ to $n(R,z,\phi)$ conversion}

We use cylinder coordinates $R, z, \phi$ to describe the Milky Way disk;
$ \phi= 0 $ is in direction $l = 0$. We use the IAU recommendations for
the galactic constants; \rsuns = 8.5 kpc and \vsuns = 220 \kms. All the
individual steps necessary for converting the observed brightness
temperature distribution $T_B(l,b,v)$ as a function of galactic
longitude, latitude, and LSR velocity to densities $n(R,z,\phi)$ have
been described before in great detail, e.g., by
\citet{Westerhout1957,Henderson1982,Burton1986,Diplas1991,
Voskes1999,Nakanishi2003} and \citet{Levine2006a}. We sketch our
procedure only briefly; but where we deviate from other authors, we
discuss those points in detail.

Previously, \citet{Levine2006a} have used the Hanning smoothed version
of the LAB survey to derive $n(R,z,\phi)$. Our approach is similar,
but we made use of the original telescope data, cleaned for
instrumental problems (the version prior to the generation of FITS
maps; \citet{Kalberla2005}). 

To derive volume densities $n(R,z,\phi)$ for each volume element
centered at $R_0,z_0,\phi_0$ we calculated the emission within this
cell. We first determined which of the observed $T_B(l,b,v)$ profiles may
show some contribution from this volume element. For each profile we
calculated column densities along the path length within the cell under
consideration. The volume density was obtained after division by the path
length. The procedure is similar to the one used by
\citet{Diplas1991}. We assumed that the gas is optically thin.  This may,
however, be incorrect for some parts of the Galactic disk at low
latitudes \citep{Burton1992}.

The LAB data are expected to predominantly consist of emission from the
Galactic \hi~ disk, gas in a roughly circular motion. But there are
contributions from the local gas, from galaxies, also from high- and
intermediate velocity clouds (HVCs and IVCs, respectively), which are not
directly connected to the disk. Different attempts have been made to
exclude such features from the data base right at the beginning of the data
processing. Some authors have interpolated and subtracted the local \hi~
emission. Automated filtering of the $T_B(l,b,v)$ data may cause unknown
biases. We decided not to filter the observational data base but rather the
$n(R,z,\phi)$ distribution. The details for our procedure are given in
Sect. 2.2.

Most important for the conversion of $T_B(l,b,v)$ to $n(R,z,\phi)$ is
the translation of velocities to distances via a rotation curve. The
exact shape of the Milky Way rotation curve is controversial, so some
assumptions are needed to overcome this deficit. Any prejudice at this
point may bias the results, so we decided to study how different
rotation curves, corresponding to different mass models, may affect the
results. We also used a generalized rotation law $v_{\rm rot}(R,z)$,
allowing the azimuthal streaming of the gas above the disk to deviate
from the rotation of the disk.

A strictly circular motion of a particle at a constant $z$ distance
above the disk is not possible. We considered the properties of a gas
{\it layer} and describe the average rotation for a volume element with
constant average volume density in a multiphase medium. We assumed that
the streaming velocity $\overline{v}_{\rm rot}$ is dominated by
gravitational forces 
\begin{equation}
\overline{v}_{\rm rot}(R,z) = \sqrt{R \ \partial \Phi(R,z) / \partial R }. 
\label{Eq_slow-rot}
\end{equation} 
Such a relation implies, for constant galactocentric radius $R$, a
decreasing rotation velocity with increasing $z$ distance. It was used
for the first time by \citet{Westerhout1957} to derive the Galactic \hi~
volume density distribution, but later publications have assumed
cylindrical rotation.

In comparing cylindrical with lagging rotation we recall the conditions
that must apply for the cylindrical case \citep[][Paper
I]{Kalberla2003}: the density distribution has to be barotropic, $p =
f(\rho)$. A barotropic fluid is an idealized fluid in which the pressure
is a function of only the density. The interstellar medium, however, is
dominated by several co-existing phases. Here we need to consider at
least two distinct \hi~ phases, a cold neutral medium, and a warm neutral
medium (CNM and WNM, respectively). These have very different
densities but are basically in pressure equilibrium \citep{Wolfire2003};
and phase transitions are possible. Barotropic conditions do not apply
to such a multi-phase medium, which probably dominates most of the
volume under consideration. It is therefore highly questionable whether
the common assumption of cylindrical rotation applies for the \hi~ gas
in the Milky Way disk. We therefore exercise the option of using both the
cylindrical and lagging rotations for our reduction. For a more detailed
discussion, see Sect. 4.1.

Except for replacing $v_{\rm rot}(R)$ by $v_{\rm rot}(R,z)$, we used the
procedure as described in \citet[][Eqs. 1\&2]{Levine2006a} including their
first order epicyclic streamline correction; which is a great
improvement over previous distance derivations and important for
overcoming discontinuities at large radial distances.
No streamline correction was applied to $R < 8.5$ kpc, since the
correction is not applicable there. Inside the solar circle the
translation between distance and velocity is ambiguous because there are
two positions along the line of sight that have the same velocity.
These positions, however, may have different $z$ distances, so their
densities may differ accordingly. 

We modeled the \hi~ distribution,
taking the expected flaring according to the mass model into
account. Incrementing along the line of sight, we calculated expected
densities at near and far distances and weighted the derived quantities
accordingly. Our method is comparable to that of \citet{Nakanishi2003},
except that they use a $sech^2$ approach according to \citet{Spitzer42}
in place of a more detailed model. We applied no correction for
densities around the tangent points, as described by
\citet[][Sect. 3.2]{Nakanishi2003} since this tends to cause
discontinuities. The focus of our paper is on data at $ R > 8.5$
kpc. However, as data inside the solar circle are found to be useful for
checking the consistency of a model, they are included in most of the
plots.

The $n(R,z,\phi)$ distribution was calculated and interpreted for $ 0 <
R < 40 $ kpc, $ -15 < z < 20 $ kpc at a resolution of $\Delta R = 100$
pc, $\Delta z = 100$ pc, and $\Delta \phi = 1 \deg$. In addition we
allowed $l, b, v$ to be traced back for each position $R, z, \phi$. To avoid
biases due to local \hi~ features we flagged and discarded all data with $|b| >
30 \deg$. Similarly we flagged data with $|v_{\rm LSR}| < 10 $ \kms. In the
direction of the Galactic center, we discarded data originating from $|R| <
3.5 $ kpc. This region, approximately with the extent of the bar, is
affected by ambiguities due to large velocity dispersions
\citep[][Fig. 8]{Weiner99}.

\subsection{Determination of the first moments}

To derive genuine disk gas we initially flagged all data for $|b| > 30
\deg$. But then we did not discard any features of the observed spectra
to distinguish between gas within the disk and gas that is, most
probably, not associated with the disk. The derived volume density
distribution contains both contributions. An inspection of the 3-D
distribution shows that it is relatively easy to distinguish between
both contributions visually, if one generates video sequences. The disk
component, in circular motion, stays mostly stationary, while there are
other components that appear to move toward the Sun as the frames
approach the Sun or, opposite, as one moves away from the Sun. Galaxies,
in particular, show up as fast-moving isolated spots. This effect is
most prominent if the video frames loop in azimuth angle $\phi$.

\begin{figure*}
   \centering
   \includegraphics[angle=-90,width=17cm]{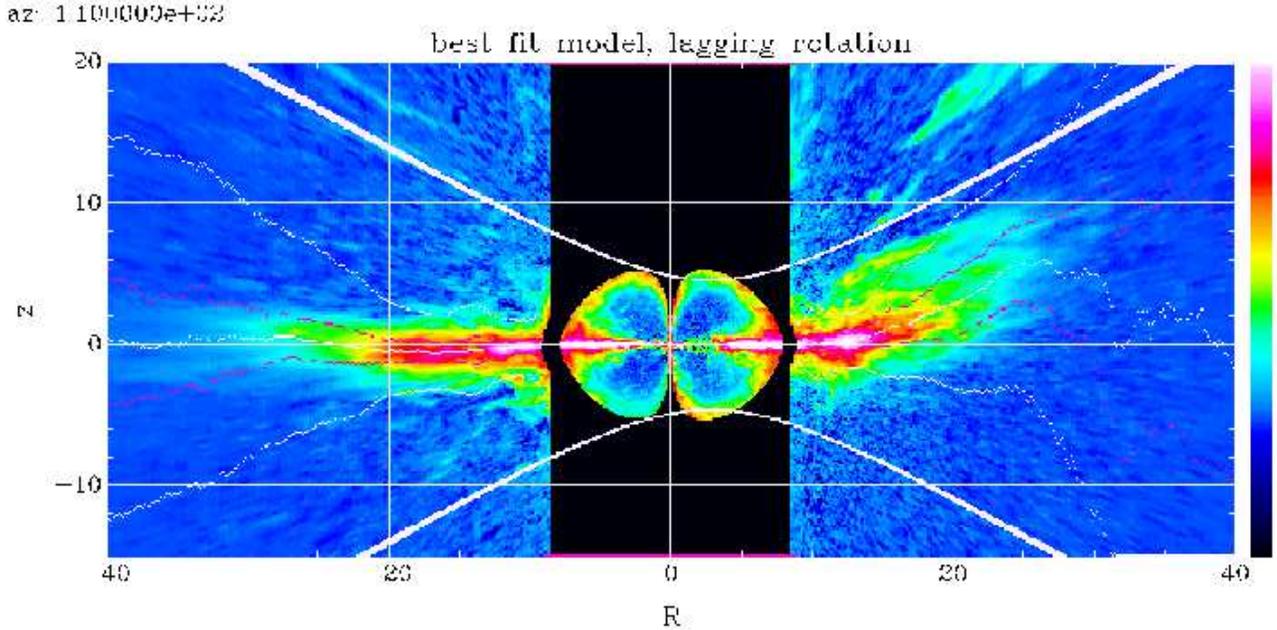}
\caption{Derived volume density distribution for our best-fit model at
  azimuth $\phi = 110 \deg$ (right hand side) and opposite, $\phi = 290
  \deg$ (left hand side) for $ 0 < R < 40 $ kpc and $ -15 < z < 20$
  kpc. Densities affected by local emission with $|v_{\rm LSR}| < 10$
  \kmss are flagged black. The curved white, thick lines indicate
  latitudes $|b| = 30 \deg \pm 0.5 \deg$. The derived mid plane of the
  \hi~ distribution is indicated by the thin white line close to $z = 0$
  kpc, and the red dots represent the dispersion of the flaring \hi~ gas
  layer. The outer white dots indicate the initial estimate for the
  extension of \hi~ gas considered to belong to the disk.  A logarithmic
  transfer function was chosen for $ n < 0.5$ cm$^{-3}$ to emphasize low
  densities.  }
         \label{Fig_az110}
  \end{figure*}

\begin{figure*}
   \centering
   \includegraphics[angle=-90,width=17cm]{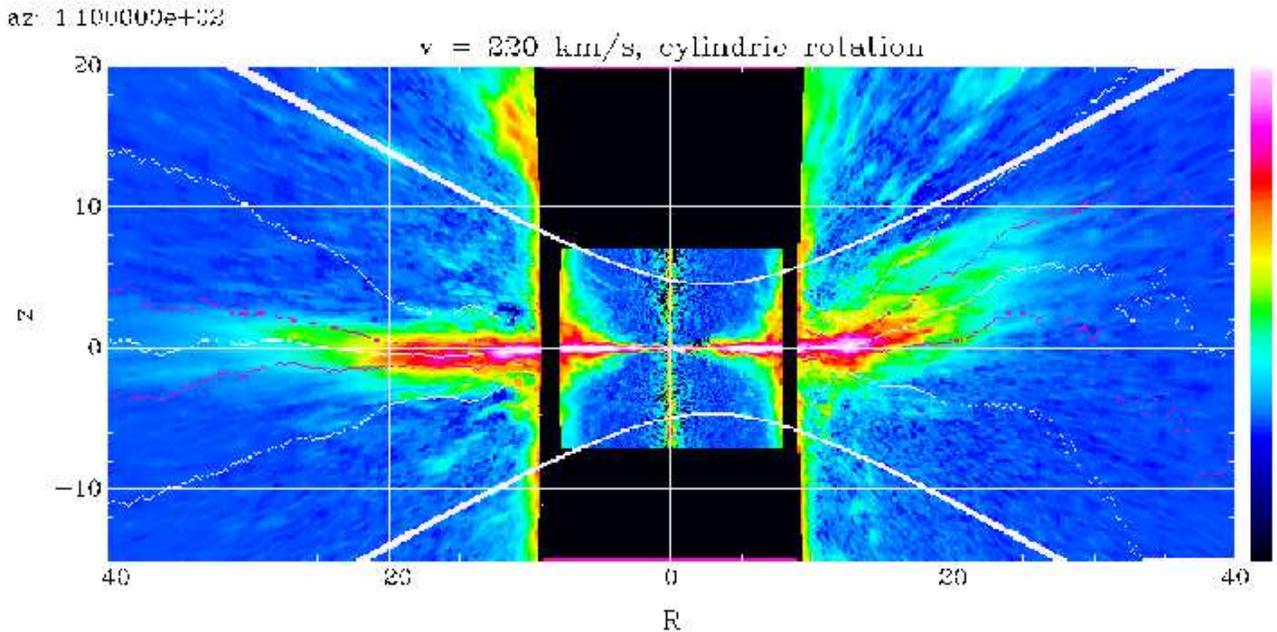}
\caption{Volume density distribution derived under the assumption 
  of a constant cylindrical rotation with $v_{\rm rot}=220$. See
  Fig. \ref{Fig_az110} for further explanation. }
         \label{Fig_az110_v220}
   \end{figure*}

To focus on the stationary disk emission, we averaged $n(R,z,\phi)$ over
15\deg \ in $\phi$. This average does not have a major effect on the
stationary disk component but degrades spurious emission caused by \hi~
gas in non-circular motion considerably. We therefore used it as an
initial guess of the first three moments; the surface density $\Sigma$,
the first moment $z_0$, representing mid-plane, and the second moment
$\sigma$, representing the scale height of the gas. As mentioned before,
data with $|b| > 30 \deg$, $|v_{\rm LSR}| < 10$ \kms, had been flagged
in the first place. In addition, we flagged all data in the original
data base outside the range $ z_0 - 2.35 \ \sigma < z < z_0 + 2.35 \
\sigma $. For a Gaussian distribution, 2\% of the data would be
discarded by this criterion. On average we find more than double that
amount. As a final step we determined the moments $\Sigma(R,\phi)$,
$z_0(R,\phi)$, and $\sigma(R,\phi)$ from those parts of the unsmoothed
data that was not flagged, thus excluding extra-planar gas.

We verified each of the reduction steps by inspecting the data cubes
visually. Most critical is the flagging process. We found the automated
procedure worked as expected for a majority of cases. However, many
features exist that provide no clear evidence whether or not they belong
to the disk. { Figures \ref{Fig_az110} \& \ref{Fig_az110_v220} show
example maps for the volume density distribution at azimuth $\phi = 110
\deg$ (right hand side) and opposite, $\phi = 290 \deg$ (left hand
side). Most of the flagging of
extra-planar gas starts already in regions below $|b| \la 30 \deg$. The
latitude limit $|b| = 30 \deg$, naively introduced to blank out local
gas at high latitudes, gets justified this way and does not affect our
analysis in a quantitative way. 

Figure \ref{Fig_az110} displays volume densities for our best-fit model,
while Fig. \ref{Fig_az110_v220} was derived from the standard model with
a constant rotation curve; for a more detailed discussion of the
differences we refer the reader to Sect. 4.1.  Prominent in Fig.  \ref{Fig_az110}
is the elongated feature at $12.5 \la R \la 16.5 $ kpc and $4.3 \la z
\la 6.3$ kpc. This is the HVC complex R located at $ 62\deg \la l \la
73\deg$ and $ 5\fdg5 \la b \la 15\deg$ with radial velocities of $ -156
\la v_{\rm LSR} \la -100$ \kms.  There are some indications that this
HVC complex may be interacting with the disk \citep{Kalberla2006}, but
this feature cannot be considered part of the stationary galactic
disk. It was removed by our automatic flagging. Similarly, some
components showing up in $T_B(l,b,v)$ as intermediate velocity clouds
(IVCs) were removed by the flagging.

Figures \ref{Fig_az110} \& \ref{Fig_az110_v220} are examples only, but
they do show clearly that warp and flaring of the \hi~ disk are
asymmetric. In particular, as discussed in Sects. 8 \& 9 in more detail,
the region $90\deg \la \phi \la 110\deg$ appears strongly perturbed for
$R \ga 15$ kpc. For this reason we exclude this azimuth range in the
following whenever we determine average properties of the \hi~
distribution. }

Finally, we emphasize that our data reduction is fully automated,
without any human interaction in the flagging process. The procedures
were identical for all the mass models and rotation curves we
considered.


\section{Mass models and volume density distributions}

Probably the most frequently used rotation law for the Milky Way
assumes a constant rotation $v_{\rm rot} (R) = 220$ \kmss
\citep{Brand1993}.  This naive model is completely free of specific
assumptions about the dark matter distribution. We use it for 
reference.

We display in Fig. \ref{Fig_az110_v220} the volume density distribution
derived from the reference model at azimuth $\phi = 110 \deg$. The most
prominent features, considered to belong to the disk, are in good
agreement with Fig. \ref{Fig_az110} for our best-fit model. It is the
changes for HVC complex R that are remarkable. This feature has almost
disappeared. The main emission from this HVC complex has shifted to an
azimuth of 104\deg, while at the same time the distance in $R$ has
increased by 2 kpc, in $z$ by 1 kpc.

\subsection{Definition of self-consistent mass models}

Our mass model contains the following components: an isothermal stellar
disk, represented by thin and thick disk components, a bar and a bulge
and isothermal gaseous disks for molecular gas, the cold neutral medium,
the warm neutral medium, diffuse ionized gas (DIG), a hot ionized, and a
cold neutral halo component. Surface densities and scale heights for all
of these components are adjusted to match the observational constraints
(we refer to \citetalias{Kalberla2003} for details).  Such a
distribution can be supplemented by any dark matter distribution that
matches known observational constraints. Most importantly it must be
consistent with the dark matter distribution on large scales
\citep{Zaritsky99}. As a constraint for isothermal dark-matter halo
models, we demand a total mass of $ \sim 2 \ 10^{12} $ \msuns within $ R
\la 350$ kpc. We demand \vsuns = 220 \kms for the local dark matter
distribution.

The mass distribution for each model is checked for self-consistency as
described in \citetalias{Kalberla2003} (it has to solve the combined
Poisson-Boltzmann equation). The visible mass distribution (stars and
gas) is allowed to adjust its scale height to the total gravitational
potential from disk and halo until convergence is achieved. The model
disk is flat, but for comparison with observations, we applied a
coordinate transformation to take warping into account
(Eq. \ref{Eqwarp}).

\subsection{Self-consistent \hi~ flaring curves}

For each mass model one has to verify whether it is consistent with the
derived \hi~ distribution. In each case we determined the rotation curve,
converted $T_B(l,b,v)$ to $n(R,z,\phi)$ and calculated the moments
$\Sigma(R,\phi)$, $z_0(R,\phi)$, and $\sigma(R,\phi)$. We then compared
the average flaring $ h(R) = f \ <\sigma(R,\phi)>$, derived from the
data, with the expected flaring. For convenience we used for the gaseous
flaring $ f = \sqrt{2 \ {\rm ln} 2} $ to describe the half width at half
maximum (HWHM) of the \hi~ layer. This convention allows an easy
comparison with the literature values.

\subsubsection{Flaring of the $v_{\rm rot}=220$ \kmss reference model}

\begin{figure}[!ht]
   \centering
   \includegraphics[angle=-90,width=9cm]{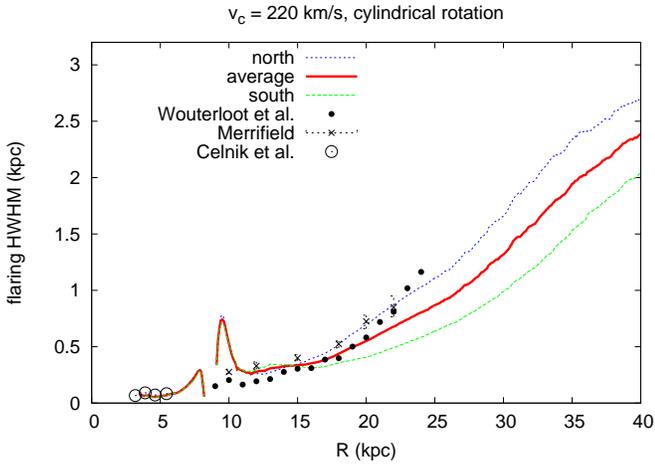}
\caption{Derived \hi~ flaring for the standard model. The thick red line
  gives the mean extension of the flaring \hi~ layer (half width at half
  maximum). The dotted blue and green lines give the flaring in the
  northern and southern hemispheres, respectively. From the literature we
  include: $\bigcirc$ derived from \citet{Celnik79}, X with error bars
  from \citet{Merrifield1992}, and $\bullet$ \citet[Table
  1]{Wouterloot90}.  }
         \label{Fig_flare_v220}
   \end{figure}

Figure \ref{Fig_flare_v220} shows the flaring curve derived for a flat,
$v_{\rm rot} = 220$ \kmss rotation curve. We excluded $90\deg \la \phi
\la 110\deg$ and those regions in direction of the center or anti-center
that might be affected by spurious emission not connected with the \hi~
disk. To allow a comparison of our flaring data with previous
determinations we include in Fig. \ref{Fig_flare_v220} some data from
the literature.  The data for $R < $ \rsuns ($\bigcirc$) were derived
from \citet{Celnik79} by averaging over ten positions from their Table
3. The literature data agree well with our results, except for $ 7\la R
\la 11$ kpc. Our analysis is insensitive within this range.

\subsubsection{Flaring caused by a spheroidal halo }

We tested several conventional halo models, some of which were 
already discussed in \citetalias{Kalberla2003}. First we discuss 
non-singular spheroids with flattening $q$ \citep[see
e.g.][]{OM01,Narayan2005},
\begin{equation}
\rho(R,z) = \rho_0 \left[\frac{R_c^2 } {R_c^2 + R^2 + (z/q)^2}\right]^p . 
\label{Eq_sphere}
\end{equation} 
Oblate halos have $q < 1$, prolate $q > 1$. $\rho_0$ is the density at
the center, $R_c$ defines the core radius, and $p$ is the power index
\citep{Narayan2005}. We consider the case $p = 1$ first.

\begin{figure}[!th]
   \centering
   \includegraphics[angle=-90,width=9cm]{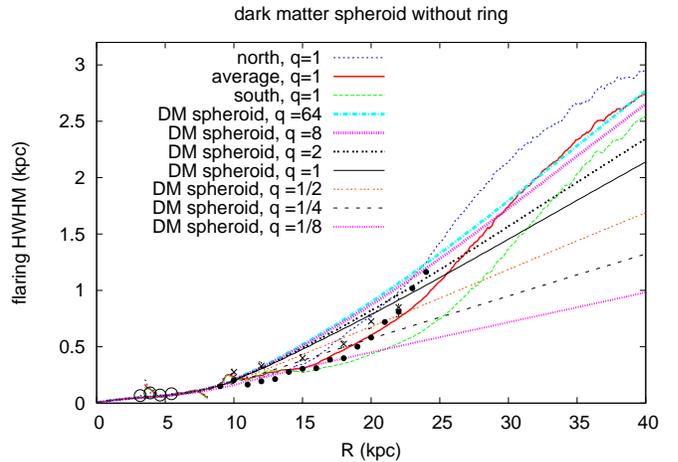}
\caption{Self-consistent \hi~ flaring for isothermal dark matter
spheroids. The thick red line gives the mean
extension of the \hi~ layer that should be comparable to the solid
black line, which represents the flaring curve expected for the model
with $q=1$. In addition, we include flaring curves for oblate halos
with axial ratios $c/a = 1/2 $ to $1/8$ and also for prolate halos
with axial ratios $c/a = 2/1 $ to $64/1$.}
         \label{Fig_flare_noDMD_noring}
   \end{figure}

Figure \ref{Fig_flare_noDMD_noring} shows the \hi~ flaring derived in a
self-consistent way from the LAB survey for a spheroidal halo with $q=1$
in comparison to flaring curves expected for mass models with various
$q$ parameters. We discuss $q=1$ first. For $ R \ga $ \rsun \ the
predicted flaring increases linearly with distance, in agreement with
the analytical solution given by \citet[][Eq. 8]{OM00}. Comparing the
model with the data, we find a systematic S-shaped deviation.  The model
curve explains the average trend but fails to reproduce details of the
flaring derived from the data around $R \sim 17$ kpc and $R \sim 35$
kpc.

To improve the model we explored how the flaring might be affected by the
flattening parameter $q$. According to \citet{OM00}, we expected changes in
slope. Oblate halos, $q < 1$, lead to a low flaring at $R \sim 17$ kpc
but fail to reproduce the observations at $R \sim 35$ kpc. Prolate halos
improve considerably for $q \ga 8$ but cannot reproduce the data at $R
\sim 17$ kpc. Each halo model also leads to a different flaring curve
when fitted to the observations. Since each model leads, iteratively, to
a different result for the rotation curve, there would be a need 
for each $q$ parameter to also have a plot for the flaring, as derived from
the data. For reasons of conciseness we refrain from showing
all. Changes are in any case less significant than the changes in the
model's flaring curves plotted in Fig. \ref{Fig_flare_noDMD_noring}.
 
Previous determinations of the flattening parameter $q$ from \hi~
flaring led to contradictory results. \citet{OM00,OM01} consider the
\hi~ flaring at $R = 2$ \rsuns and advocate $q = 0.8$. They conclude
that significantly flattened halos are possible only in the case that
the distance of the Sun from the Galactic center is smaller than 6.8
kpc.  \citet{Narayan2005} used data in the range $8 \la R \la 24 \la$
kpc to find that isothermal spheroids do not fit the observed flaring data
particularly well. Using only the flaring at $R = 2$ \rsuns, we could
conclude that the halo must be rather oblate, $q \sim 0.1$, but
Fig. \ref{Fig_flare_noDMD_noring} shows that the halo model needs to
become increasingly prolate with distance. This reinforces the
conclusion of \citet{Narayan2005} that a halo described by a single
constant $q$ parameter does not match the observations if one considers
a broad range of galactocentric distances.

\subsubsection{Flaring caused by a dark matter disk}

\begin{figure}[!ht]
   \centering
   \includegraphics[angle=-90,width=9cm]{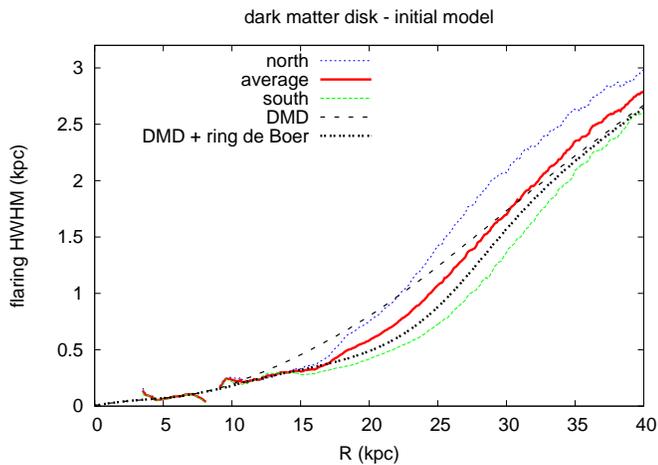}
\caption{Derived \hi~ flaring for the initial dark matter disk model. We
  include the flaring curve as expected from the initial dark matter
  disk model (\citetalias{Kalberla2003}, black dashed) and after adding
  a dark matter ring to the model with parameters as given by
  \citet{deBoer2005} (black dotted). }
         \label{Fig_flare_ini}
   \end{figure}

We tested alternatives to standard spheroidal halos and found that it is
much easier to match a model to the observations if one considers dark
matter associated with the Galactic disk. An example for the derived
flaring curves is given in Fig. \ref{Fig_flare_ini}, with what we expect
from our initial dark matter disk model (DMD)
(\citetalias{Kalberla2003}). For comparison we plot the mean flaring
determined from the derived $n(R,z,\phi)$ distribution for
the northern and southern parts of the Milky Way disk.

Comparing the flaring derived from observations with the model, we
generally find reasonable agreement between the two, comparable to the
highly prolate spheroidal mass model ($q \sim 8$) from
Fig. \ref{Fig_flare_noDMD_noring} but with a better match for $ 12 < R <
25$ kpc. However, we still have significant S-shaped deviations.
We investigated whether these deviations could be explained by reduction
problems. Each of our attempts to explain the discrepancies with
problems during the data processing failed. Even trying deliberately to
bias the data reduction led to no significant changes for the
resulting \hi~ distribution. We conclude that the deviations between
model and data are highly significant and not caused by a systematic
bias.

The depression in the flaring curve at $ 12 < R < 25$ kpc has been seen
before, see e.g. the recent papers by \citet{Nakanishi2003} and
\citet{Levine2006a}, but the effect shows up as most pronounced in Fig. 23
of \citet{Voskes1999}.  A low flaring can be explained either by an
extraordinary low velocity dispersion of the \hi~ gas or by a local
enhancement of the mass distribution. The turbulent velocity dispersion
in the interstellar medium (ISM) appears to be independent of distance
\citep{Blitz1991,Burton1992}.  We find therefore no reason to assume
that the \hi~ gas is special at distances $15 \la R \la 20$ kpc.

\subsubsection{Adding a dark matter ring }

None of the well-established dark matter models has an enhancement of
the mass distribution at a distinct radial distance from the center,
something our models seem to require to reduce or even eliminate the
S-shape deviation shown in Fig. \ref{Fig_flare_ini}.  However, a
possible scenario could be a ring-like mass concentration caused by a
merger event; see \citet{Hayashi2003} and
\citet{Penarrubia2005,Penarrubia2006}. Several groups claimed to have
detected stellar streams in the Milky Way, right at the region of
interest:
\citet{Newberg2002,Ibata2003,Yanny2003,Martin2005,Martinez2005,Bellazzini2006}.
Others explain the over-densities as due to the Galactic warp:
\citet{Momany2004,Momany2006} and \citet{Lopez2006}. The stellar mass of
the giant stellar structure is estimated to be $ 2 \ 10^{8}$
\msuns--$10^9$ \msuns \citep{Ibata2003}. It is, however, readily shown
that such a mass is insufficient for explaining the low flaring at $R
\sim 18$ kpc.

\begin{figure}[!ht]
   \centering
   \includegraphics[angle=-90,width=9cm]{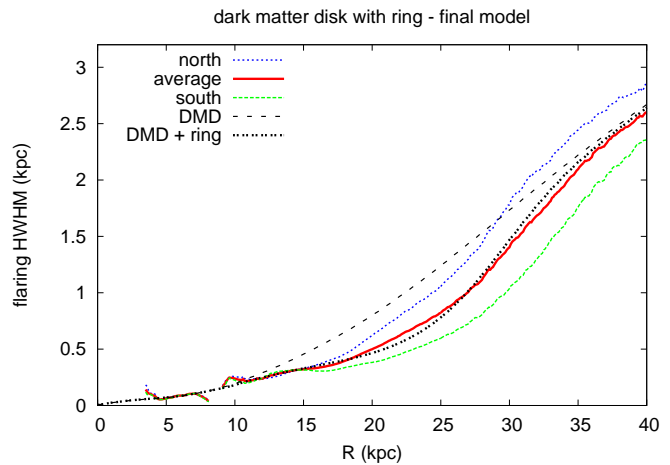}
\caption{Observed mean HWHM flaring (thick red line), applying the
 rotation curve of our best-fit model.  The dotted blue and green lines
 give the flaring in the northern and southern hemispheres, respectively.
 The best-fit flaring model, containing dark matter disk and ring, is
 plotted with a black dotted line. The flaring curve from the initial
 dark matter disk model (without ring) is shown for comparison (black
 dashed).  }
         \label{Fig_flare_final}
   \end{figure}

A ring with a mass content well above the estimates for the stellar mass
contribution is needed, and exactly this was recently claimed by
\citet{deBoer2005} to exist. They studied the diffuse Galactic
$\gamma$-ray background observed with EGRET. They find excess emission
close to the Galactic disk and claim that the excess originates from
ring-like dark matter distributions. One of the rings is located at $R
\sim 14$ kpc.

To test the \citet{deBoer2005} proposal, we include the ring at $R \sim
14$ kpc in our model (Fig. \ref{Fig_flare_ini}). The additional dark
matter ring indeed provides a step in the right direction, but obviously
the predicted flaring is now too low. We varied the position of the
ring, the radial dispersion, and its total mass. All calculations
included a self-consistent solution of the Poisson-Boltzmann equation, a
derivation of the Galactic velocity field, and finally a redetermination
of the 3-D volume density distribution $n(R,z,\phi)$ and the resulting
flaring.

The best result is shown in Fig. \ref{Fig_flare_final}. Note that not
only the model flaring curve, but also the flaring derived from the data
has changed in comparison to Figs. \ref{Fig_flare_v220} to
\ref{Fig_flare_ini}.  The model reproduces the observations for a
ring-like mass distribution
\begin{equation}
n(R,z) = n_r~ e^{-(R - R_r)/(2 \sigma_r^2) - |z/z_r|}
\label{Eq_ring}
\end{equation} 
with $n_r = 0.005$ cm$^{-3}$, $ R_r = 17.5$ kpc, $\sigma_r = 5$ kpc, and
$z_r = 1.7$ kpc. This relation was given by \citet{deBoer2005}, but all
parameters, except $z_r = 1.7$ kpc, have been redetermined and modified
by us for an axisymmetric model. { Discrepancies between our solution
and parameters given by \citet{deBoer2005} may be explainable, at least
partly, by uncertainties and systematical deviations of the ring-like
mass distribution from axisymmetry. We determined here the {\it mean}
properties of the ring alone. In Sects. 8 \& 9 we discuss
fluctuations in more detail. The method used by \citet{deBoer2005} is
most sensitive in direction to the anti-center, a region which is
seriously affected by velocity crowding, therefore not accessible to our
analysis. }

The resulting ring has a total mass of $2.3~ 10^{10}$ \msun.  For the
remaining part of the dark matter model we kept the model parameters as
described in Table 1 of \citetalias{Kalberla2003} with only two
changes. The local dark matter disk density had to be adjusted from $ n
=0.69$ cm$^{-3}$ to $ n = 0.75$ cm$^{-3}$, and we adopted a radial scale
length $H_R/2 = 3.75$ kpc for a better fit of the \hi~ gas surface
density.

\subsubsection{Spheroidal halo with dark matter ring }

\begin{figure}[!th]
   \centering
   \includegraphics[angle=-90,width=9cm]{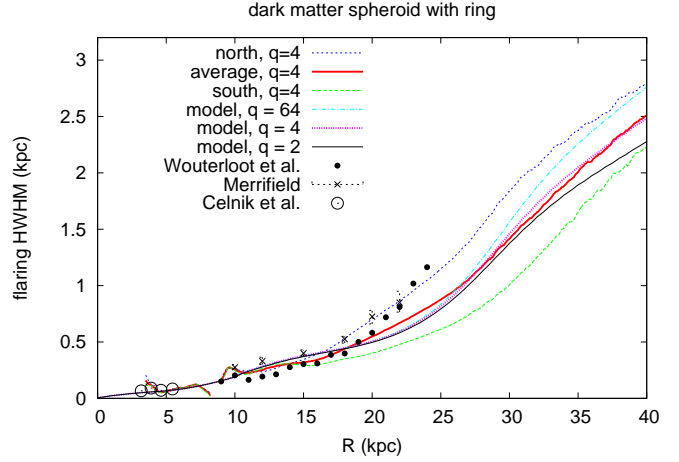}
\caption{Derived \hi~ flaring for an isothermal dark matter spheroid
with a dark matter ring. The symbols are the same as in Fig.
\ref{Fig_flare_final}. For comparison, flaring models are given
for $ q = 2 $, 4, and 64. }
         \label{Fig_flare_noDMD}
   \end{figure}

Next we tested whether the inclusion of a dark matter ring can also lead
spheroids to fit better. { We consider three cases, $ q = $2, 4, and
64. Figure \ref{Fig_flare_noDMD} shows that a prolate spheroid with $ q
= 4$ represents a good fit, but observations cover the range $q = 64$
for the northern and $q = 2$ for the southern part of the Milky Way. The
ring parameters used are identical to those discussed in Sect. 3.2.4,
except that we need only 80\% of the total
ring mass in the case of a spheroidal halo.  Solutions with different flattening parameters $q$ are
nearly identical for $R \la 25$ kpc. This emphasizes once more the need
to get constraints at large distances. Milky-Way mass models are
ambiguous if a limited data base is used, a point stressed previously by
\citet{Dehnen98}.

Comparing Fig. \ref{Fig_flare_noDMD} with Fig.  \ref{Fig_flare_final}
formally results in an acceptable solution for $q = 4$. Further
improvements to the model (e.g. at $R \sim 15$ kpc and $R \sim 22$ kpc)
would be possible by splitting the dark matter ring into several
lumps. We did not explore them for two reasons. i) Strongly
prolate spheroids are rather unexpected from cold dark matter
simulations \citep[e.g.][]{Dubinski1991}, and ii) the surface densities
at $R \sim $ \rsuns are unacceptable for all of our spheroidal models.
In particular, highly prolate spheroids show the most significant
deviations from HIPPARCOS results, see to Sect. 6 for further
discussion.}

\subsubsection{NFW halo with ring}

\begin{figure}[!th]
   \centering
   \includegraphics[angle=-90,width=9cm]{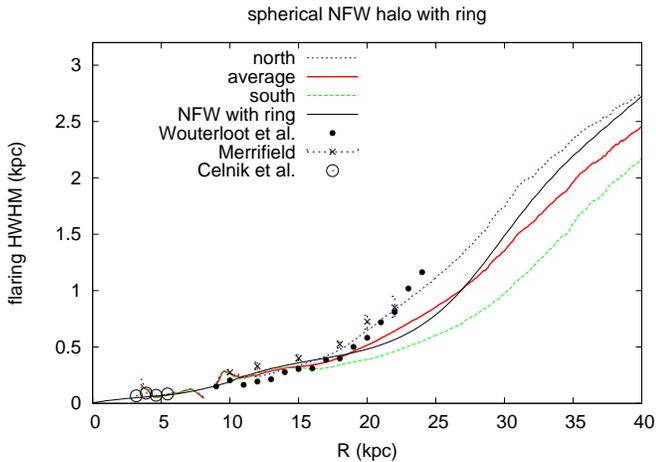}
\caption{Derived \hi~ flaring for the NFW model. The solid black line
  gives the model. See Fig. \ref{Fig_flare_noDMD} for further
  explanation.}
         \label{Fig_flare_NFW}
   \end{figure}

We also tried to fit a $\Lambda$CDM-based NFW model \citep{NFW}, using
parameters provided by \citet{Klypin2002}. The result is shown in Fig.
\ref{Fig_flare_NFW} and fits the observations as well as the results
presented in Figs. \ref{Fig_flare_final} \& \ref{Fig_flare_noDMD} and
might be improved by assuming a lumpy ring structure. For a favored halo
concentration parameter $C = 12 $, we needed, however, to modify the
central mass density to keep $ v_{\rm rot} = 220$ \kms, resulting in a
total halo mass of $ 5\ 10^{11}$ \msun.  This is in clear conflict with
our constraint that the total mass of the Milky Way within $ R \la 350$
kpc should be close to $ 2 \ 10^{12} $ \msun.  Moreover, the halo mass
is only 50\% of the solution demanded by \citet{Klypin2002}, likewise in
conflict with the $\Lambda$CDM constraints and therefore not
acceptable. We were unable to match the flaring at large distances
exactly with the observations (Fig.  \ref{Fig_flare_NFW}). It would be
necessary to increase the halo mass, but this would imply at the same
time that the local circular speed must significantly exceed $v_{\rm
rot} = 220$ \kmss. For the NFW model considered, there is too much mass
in the inner part of the Galaxy. This is the well-known cusp problem.

\section{Galactic rotation}

Each of the mass distributions discussed in the previous section
causes a different rotation curve. This was taken into account when we
derived self-consistent flaring curves for these models. Here we restrict
the discussion to the dark matter disk model.

\begin{figure}[!ht]
   \centering
   \includegraphics[angle=-90,width=9cm]{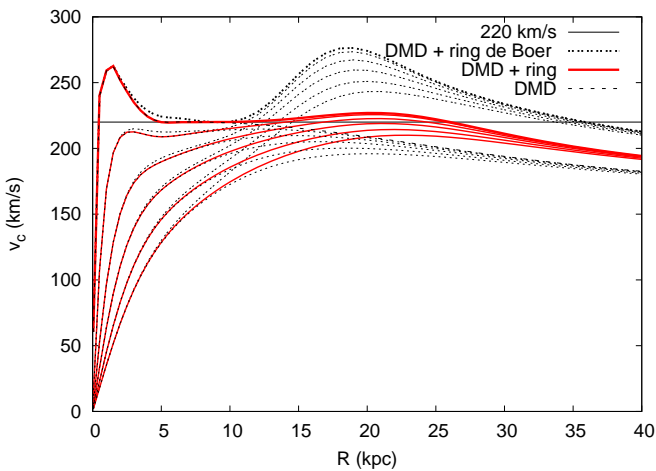}
\caption{Rotation curves for the models displayed in
  Figs. \ref{Fig_flare_ini} \& \ref{Fig_flare_final}. The black dashed
  lines give the rotation curves for the initial dark matter disk, and the
  dotted lines the curves for the dark matter disk including a ring
  according to \citet{deBoer2005}.  The thick red lines show the
  rotation curve for the best-fit model. To the rotation at $ z = 0$ kpc
  (top thick curves) we add in each case the circular streaming
  velocities at $ |z| = 1$ to 5 kpc (from top to bottom).  }
         \label{Fig_rot_curve}
   \end{figure}

The 3-D rotation curves for this model are shown in
Fig. \ref{Fig_rot_curve}. We plot the initial dark matter disk, the proposal by \citet{deBoer2005}, and the
final best-fit. In each case we show the
circular streaming velocities at $ |z| = 0$ to 5 kpc (from top to
bottom). Including a dark matter ring in our model causes the initially
falling rotation curve of the dark matter disk (DMD) model to rise
significantly at $ R \sim 18$ kpc. Adjusting the mass density of the
ring to match the \hi~ flaring brings the rotation curve down. Our best-fit mass model leads to a surprisingly simple result. The rotation law
at $z = 0$ kpc is almost flat over a range $5 \la R \la 27 $ kpc.

A flat rotation curve appears to conflict with observations
\citep[e.g.][]{Blitz1979,Schneider1983,Fich1989,Merrifield1992,Brand1993,Honma1997}
which instead result in the high velocities of the \citet{deBoer2005}
model. However, the discrepancy can be solved immediately if one takes
distance uncertainties into account. The tracers used by various authors
to measure the Galactic rotation curve for $R \ga 11 $ kpc have large
distance errors. \citet{Binney1997} assumed that most of these objects
are located within a ring at $R \sim 14$ kpc and derived for this case a
flat rotation curve.

\subsection{The co-rotation problem}

\begin{figure}[!th]
   \centering \includegraphics[angle=-90,width=9cm]{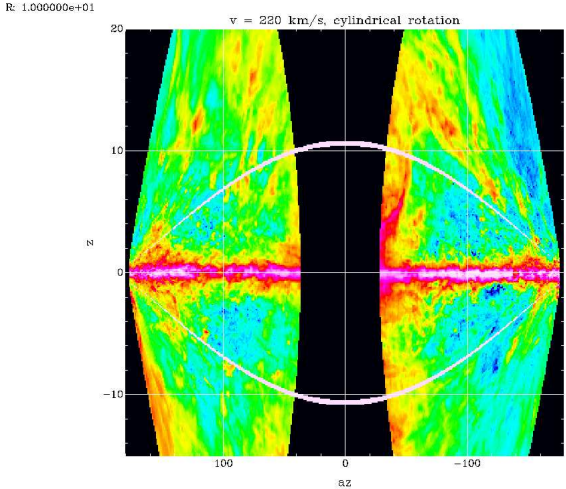}
\caption{Volume density distribution of the standard model, assuming
  cylindrical rotation, derived at $ R = 10$ kpc for azimuth $ 180\deg >
  \phi > -180\deg$ and $ -15 < z < 20 $ kpc.  Densities affected by
  local emission with $|v_{\rm LSR}| < 10$ \kmss are flagged dark.  The
  white lines indicate latitudes $|b| = 30 \deg \pm 0.5 \deg$. Only
  densities below $ n < 0.5$ cm$^{-3}$ are displayed. We use the same
  logarithmic transfer function as in Figs. \ref{Fig_az110} \&
  \ref{Fig_az110_v220}. }
         \label{Fig_R10_0}
   \end{figure}

\begin{figure}[!h]
   \centering \includegraphics[angle=-90,width=9cm]{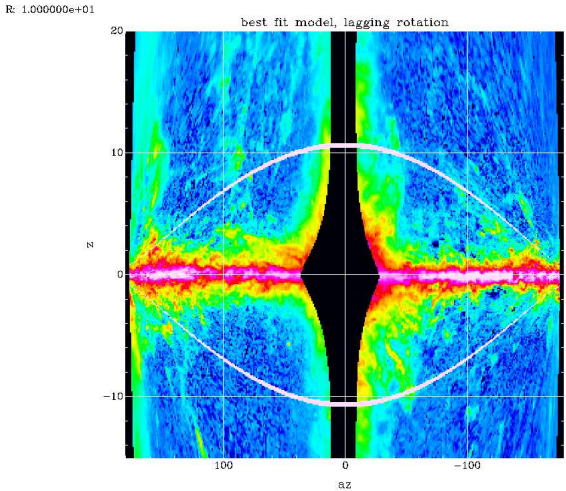}
\caption{Volume density distribution derived at $ R = 10$ kpc for the
best-fit model with a lagging rotation according Eq. 1. For more 
details see Fig. \ref{Fig_R10_0}.  }
         \label{Fig_R10_8}
   \end{figure}

In Fig. \ref{Fig_R10_0} we display the volume density distribution at
a constant galactocentric distance $R = 10$ kpc, derived for the
standard model with a constant rotation velocity $ v_{\rm rot} = 220$
\kms. Figure \ref{Fig_R10_8} displays the distribution at the same
distance, but for our best-fit model. Discrepancies are obvious, because
they
are caused by the different rotation curves. The velocities at $z = 0$
kpc are identical for both cases (see Fig. \ref{Fig_rot_curve}).
Correspondingly, the densities at $z = 0$ kpc are the same.  The
standard model (Fig. \ref{Fig_R10_0}) has a cylindrical rotation, but
for the best-fit model (Fig. \ref{Fig_R10_8}) $v_{\rm rot}(R,z)$ lags
according to Eq. 1 for increasing $z$ distances increasingly behind the
disk.

We emphasize that Figs. \ref{Fig_R10_0} \& \ref{Fig_R10_8} have
been derived by running identical procedures, the only difference being the
3-D velocity field. Comparing distinct features off the plane in the 3-D
volume density distribution $n(R,z,\phi)$ for both rotation laws, we
find that structures, most noticeabley cloud complexes, are shifted to
larger $R$ and $|z|$ distances if one assumes a cylindrical rotation. An
example, the HVC complex R, was discussed in Sect. 3.1. 

\begin{figure}[!th]
   \centering
   \includegraphics[angle=-90,width=9cm]{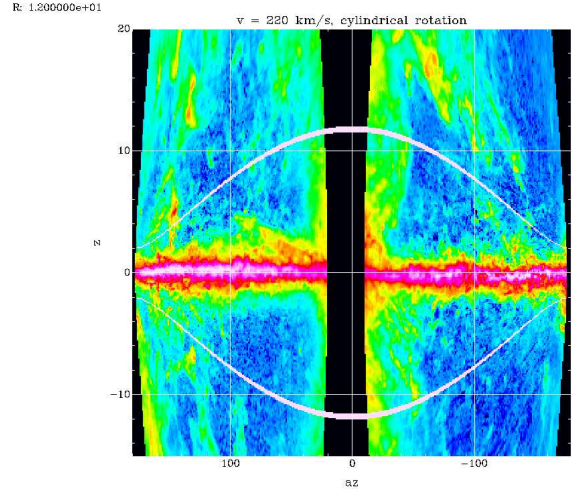}
\caption{Volume density distribution derived for $ R = 12$ kpc for the
standard model with cylindrical rotation. }
         \label{Fig_R12_0}
   \end{figure}

\begin{figure}[!ht]
   \centering
   \includegraphics[angle=-90,width=9cm]{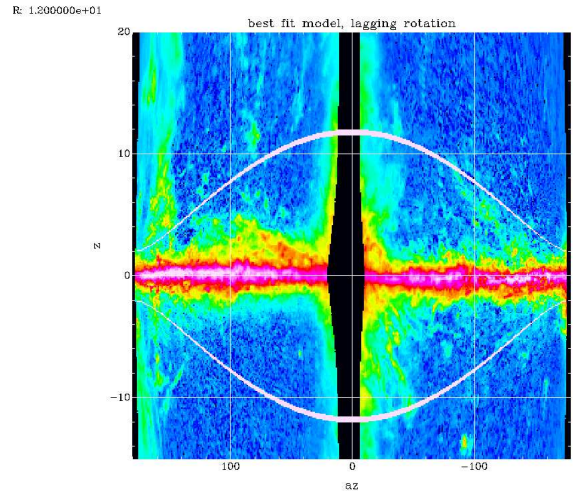}
\caption{Volume density distribution derived for $ R = 12$ kpc for the
best-fit model with a lagging rotation law. }
         \label{Fig_R12_8}
   \end{figure}

The average \hi~ volume density of the disk $<n(R,z)>$ falls off
exponentially with radius $R$ and height $|z|$. Shifting distinct
features to larger $R$ and $|z|$ distances therefore leads on average to
significantly higher densities for $|z| > 0$ kpc.  This effect is
obvious if one compares the volume density distribution at a constant
distance $R$ as derived for cylindrical rotation with that derived from
a lagging rotation model. The flaring is larger if one assumes
cylindrical rotation (Figs. \ref{Fig_flare_v220} and
\ref{Fig_flare_final}). Discrepancies are most significant close to the
solar circle, but Figs. \ref{Fig_R12_0} and \ref{Fig_R12_8} derived at
$R = 12$ kpc demonstrate that even larger distances may be affected.
 
For our analysis we chose the lagging rotation according to
Eq. \ref{Eq_slow-rot} since it considerably improves the internal
consistency of our results in comparison to a cylindrical rotation.  We
neglect buoyancy and any dependency of gas pressure terms $\nabla p /
\rho$ on radius \citep[][Sect. 5.5.1]{Kalberla2003}. These second-order
corrections, discussed also by \citet{Benjamin2002}, may get important
for extra-planar gas layers but are not expected to be significant for
the disk gas. For example, at $R = 20$ kpc, we expect a 10\% effect from
the $\nabla p / \rho$ term only for $z \ga 14$ kpc. This $z$ distance is
an order of magnitude off from the scale height of the disk gas so we
ignore this correction for computational reasons.

Our preference for a lagging rotation with increasing $z$ distance is
supported by observations that probe extra-planar gas, located on top of
the standard disk gas we consider here. The average rotational streaming
velocity in the Milky Way, as derived from interstellar absorption lines
at $z$ distances up to a few kpc, is found to lag behind the disk
\citep{deBoer1983,Savage1997}. \citet{Rand97,Rand00} and
\citet{Heald2006a,Heald2006b} find evidence of lagging ionized gas in
the halos of NGC 891 \& NGC 5775.  Very sensitive \hi~ maps of NGC 891
\citep{Swaters1997,Fraternali2005} and NGC 2403
\citep{Fraternali2001,Fraternali2004} similarly support a lagging
extra-planar \hi~ gas phase.

\section{Consistency check - available flaring data }

\subsection{Gaseous flaring inside the solar circle}

\begin{figure}[!h]
   \centering
   \includegraphics[angle=-90,width=9cm]{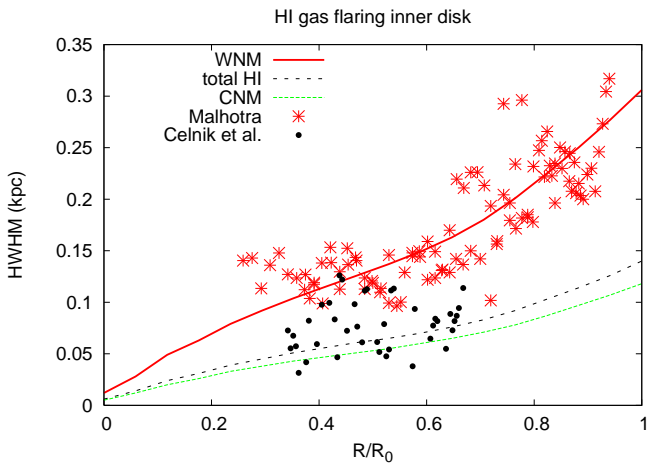}
\caption{Flaring of the \hi~ gas in the inner Galaxy. The data,
  \citet{Malhotra95}, ($\ast$), and \cite{Celnik79}, ($\bullet$),
  represent the observed HWHM of \hi~ clouds.  The dashed black line
  represents the HWHM as derived for a multiphase \hi~ medium, the
  green dashed line is for the CNM, and the red solid line 
  represents the WNM.}
         \label{Fig_flare_HI_inner}
   \end{figure}

Our approach does not allow an accurate determination of the flaring
inside the Solar circle, but we may compare our derived flaring with more
accurate data. Figure \ref{Fig_flare_HI_inner} shows the flaring for the
\hi~ gas derived by \citet{Celnik79} ($\bullet$) and by
\citet{Malhotra95} ($\ast$). The results from these groups are found to
be largely discrepant.

The scale height, derived by us from the model, is
the scale height for a mix of the CNM and WNM. Both phases have significantly different scale heights.
\citet{Celnik79} have shown that the \hi~ volume density distribution
perpendicular to the Galactic plane can be described by a
sech$^{\alpha}(z/z_0)$ law.  This relation can be
approximated by a Gaussian distribution exp[$-z^2/(2 z^2_g)$] close to
the plane and at
large distances by an exponential, exp$(-z/z_e)$. \citet{Bahcall84a}
demonstrated that such an approximation also applies for a
multicomponent model if the gravitational potential can be approximated
by a single dominating mass component. This approximation remains valid
in the case of a more general mass distribution (see
\citetalias[][Sect. 5.2.1]{Kalberla2003}). \citet{Celnik79} fitted a
general sech$^{\alpha}(z/z_0)$ volume density law, solving for
$z_g$ and $z_e$. This is the most accurate and unbiased way to determine
the \hi~ scale height. The determination of the first moment by us,
excluding \hi~ gas extending to large $z$ distances, is also
unbiased. The approach by \citet{Malhotra95} is quite different. She
assumed that the \hi~ layer can be fitted by a {\it single} Gaussian
component, represented only by the WNM (Fig. \ref{Fig_flare_HI_inner}).
Taking the different approaches into account, the flaring of the \hi~ gas
within the Solar circle is consistent with our model.

\subsection{Molecular gas}

\begin{figure}[!th]
   \centering
   \includegraphics[angle=-90,width=9cm]{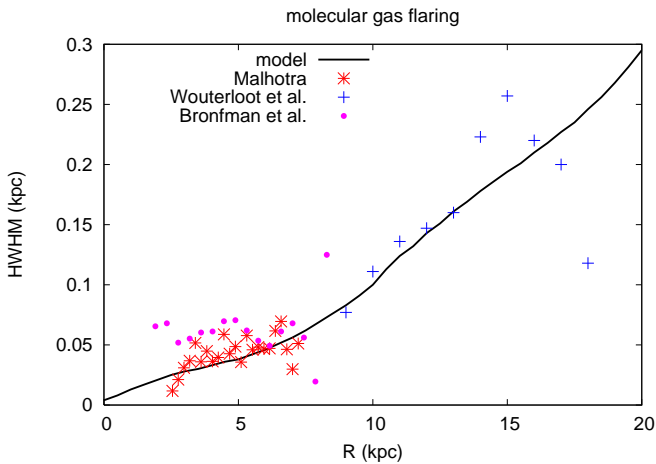}
\caption{Flaring of the molecular disk. The solid line represents the
  flaring for the final model. The data points give CO scale heights as
  derived by \citet{Malhotra94}, ($\ast$), \citet{Wouterloot90}, (+) and
  \cite{Bronfman88}, ($\bullet$). }
         \label{Fig_flareCO}
   \end{figure}

In Fig. \ref{Fig_flareCO} we compare scale heights for the molecular gas
phase from the best-fit dark matter disk model with observed scale
heights.  In the inner Galaxy two independent analyses are available 
from \citet{Bronfman88} and \citet{Malhotra94}. Both sets of values
appear to be slightly offset from each other. This may be explainable by
details of the data analysis but is certainly unaffected by
complications from a multi-phase composition as discussed in the
previous section. For the outer Galaxy we used scale heights as derived
by \citet{Wouterloot90} for a verification of the model. In general we
find good agreement between model and data except for the outermost
data points.  At distances $ R \ga 15$ kpc, the observations may be limited
in sensitivity.

\section{Consistency check - local mass surface densities}

Among the various observational constraints that may be used for 
determining the dark matter properties, the surface density
$\Sigma_{1.1}$ within 1.1 kpc is the measure with the lowest relative
uncertainties \citep{KG91}. It therefore provides one of the most
important constraints for models. Recent surface-density determinations
based on HIPPARCOS have added further constraints. 

{ \citet{KG91} determined $\Sigma_{1.1} = (71 \pm 6)\ {\rm M_{\sun}
pc^{-2}}$ using K dwarfs.  \citet{HF2004} analyzed HIPPARCOS K giants
and find $\Sigma_{1.1} = (74 \pm 6)\ {\rm M_{\sun} pc^{-2}}$. Our
best-fit model provides a consistent solution for $\Sigma_{1.1} = (79
\pm 2)\ {\rm M_{\sun} pc^{-2}}$. Within 800 pc \citet{HF2004} determine
a surface density of $\Sigma_{.8} = (65 \pm 6)\ {\rm M_{\sun} pc^{-2}}$,
we get $\Sigma_{.8} = (66.9 \pm 2)\ {\rm M_{\sun} pc^{-2}}$.  For
distances up to 350 pc, \citet{HF2004} determine $\Sigma_{.35} = 41\ {\rm
M_{\sun} pc^{-2}}$, and \citet{Korchagin2003} find $\Sigma_{.35} = (42 \pm
6)\ {\rm M_{\sun} pc^{-2}}$. Our result is $\Sigma_{.35} = (42.7 \pm 2)\
{\rm M_{\sun} pc^{-2}}$.

For $z = 0 $ pc, we find an Oort limit of $\rho_{\sun} = 0.097\ {\rm
M_{\sun}pc^{-3}}$ in excellent agreement with \citet{HF2000}, who
obtained the same value, and only marginally deviating from $\rho_{\sun}
= 0.10 \ {\rm M_{\sun}pc^{-3}}$ derived by \citet{KG89b} and
$\rho_{\sun} = (0.105 \pm 0.005)\ {\rm M_{\sun}pc^{-3}}$ by
\citet{Korchagin2003}.  The total surface density for all disk material
(stars and gas) in our best-fit model is $52.5\ {\rm M_{\sun}pc^{-2}}$,
almost identical with the model estimate $52.8\ {\rm M_{\sun}pc^{-2}}$
by \citet{HF2004}. Summarizing, the local volume and surface densities
are in excellent agreement with constraints by \citet{KG91} and with all
constraints derived from HIPPARCOS. The uncertainties given for our
model are based on typical parameter changes corresponding to
modifications of the ring parameters within an acceptable range. Our
initial dark matter disk model without ring had $\Sigma_{1.1} = 72.5\
{\rm M_{\sun}pc^{-2}}$, but adding the dark matter ring parameters as
initially proposed by \citet{deBoer2005} resulted in $\Sigma_{1.1} =
88.3\ {\rm M_{\sun}pc^{-2}}$, too high in comparison to the HIPPARCOS
result of $\Sigma_{1.1} = (74 \pm 6)\ {\rm M_{\sun} pc^{-2}}$. This
model got acceptable only after tuning with respect to the observed \hi~
flaring (Sect. 3.2.4).

All of the conventional halo-mass models, discussed in
Sects. 3.2.2, 3.2.5 and 3.2.6 have surface densities $\Sigma_{1.1}$ that
deviate by at least 30\% from $\Sigma_{1.1} = (74 \pm 6)\ {\rm M_{\sun}
pc^{-2}}$, derived from HIPPARCOS \citep{HF2004}. Extreme prolate
spheroids (Figs. \ref{Fig_flare_noDMD_noring} and \ref{Fig_flare_noDMD})
show the worst deviations. }

\section{The Galactic warp}

It is well known since \citet{Westerhout1957} that the Milky Way \hi~
disk is warped. We need to take this into account. The gravitational
potential for our model is calculated for a flat system, but we apply a
geometrical transformation afterwards to warp the system. We describe
$W(R,\phi)$ as the deviation of the center of the gas layer from $z=0$
by
\begin{equation}
W(R,\phi)=W_0(R) + W_1(R) {\rm sin}(\phi-\phi_1(R)) + W_2(R) {\rm sin}(2 \phi-\phi_2(R)),
\label{Eqwarp}
\end{equation} 
where $W_i$ and $\phi_i$ the amplitudes and phases of mode $i$.

\begin{figure}[!ht]
   \centering
   \includegraphics[angle=-90,width=9cm]{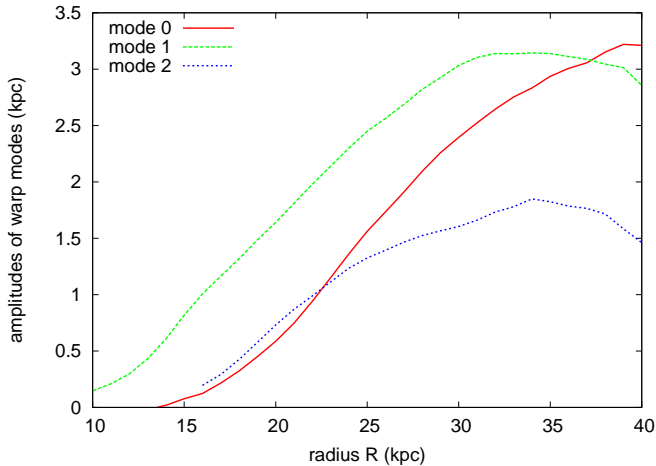}
\caption{Amplitudes for the warp modes 0 to 2 derived for the best-fit
model. Mode 2 is insignificant for $ R < 15.5 $  kpc.}
         \label{Fig_warp_amp}
   \end{figure}

\begin{figure}[!ht]
   \centering
   \includegraphics[angle=-90,width=9cm]{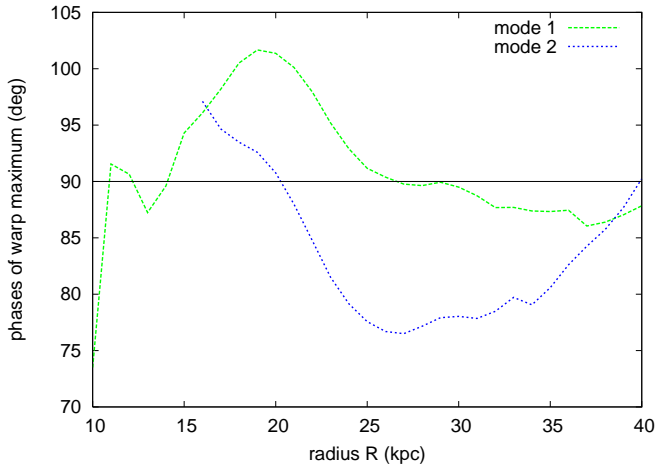}
\caption{Phases for the warp modes 1 and 2 derived for the best-fit
model. Mode 2 is insignificant for $ R < 15.5 $  kpc.}
         \label{Fig_warp_phase}
   \end{figure}

To derive the warp we averaged the observed densities within rings having
a width of $\Delta R = 1$ kpc for $R$ = 10 to 40 kpc. For each annulus
we fit the modes 0 to 2 simultaneously using a least square sine wave
fit based on the Marquard/Levenberg algorithm \citep{Press1986}. Mode 2
is found to be insignificant for $R < 16$ kpc. The fit in this range is
therefore without mode 2. We found that azimuth angles $ 80\deg \la \phi
\la 130\deg$ for $R \ga 16$ kpc show significant perturbations and
exclude this range \citep[region X as discussed by][]{Levine2006a}.

The formal errors for our fitted parameters are quite low, a few percent
for the amplitudes, and typically less than a degree for the
phases. Only at large distances do the errors increase for mode 2.
At $R \sim 40$ kpc we find a 10\% amplitude error, the typical mode 2
phase error is 3\deg.  

Figure \ref{Fig_warp_amp} displays the amplitudes for the basic warp
modes. We find a good agreement with \citet{Levine2006a}. In
Fig. \ref{Fig_warp_phase} we display the warp phases. We followed the
convention used by \citet{Levine2006a} and plot the phases for the first
maximum close to azimuth 90\deg. Here we find some differences for the
phase of mode 2, most probably due to the different rotation curve used
by us. In Fig. \ref{Fig_warp_phase} we see a well-defined maximum in the
phase of mode 1 between $ 15 \la R \la 25$ kpc. This appears to be
associated with an almost constant slope in the phase of mode 2. Formal
errors for the phases in this range are very low, less than one degree.

The warp modes are well-defined. The amplitude of mode 0 rises
continuously up to $R \sim 40$ kpc. Modes 1 and 2 both show a comparable
trend in their amplitudes, peaking close to 33 kpc. The phases are also
quite regular.

\section{Deviations from axisymmetry} 

We modeled the flaring of the Milky Way \hi~ gas assuming that the
distribution of mass and gas in the Milky Way is in equilibrium. An
axisymmetric model was used to fit the mean flaring. Figures
\ref{Fig_flare_v220} to \ref{Fig_flare_NFW} indicate systematic
differences for azimuth angles $\phi \la 180\deg$ (northern sky) and
$\phi \ga 180\deg$ (southern sky), and all flaring curves show in general
the same trends: The axisymmetric model does not reproduce the flaring
{\it exactly} but represents {\it average} properties of the mass
distribution, implying a massive exponential disk with an additional
massive ring at $R \sim 17.5$ kpc. Here, we intend to discuss systematical
perturbations from axisymmetry.

{

\subsection{A lopsided disk}

We calculated the average flaring within 20\deg \ wide sectors. Figure
\ref{Fig_flare_south} shows for the southern part of the Galactic disk
that the sector averages are well-defined, deviating only a few percent
from the mean. In the northern part (Fig. \ref{Fig_flare_north}) the
situation is very different. For radial distances $R \ga 25$ kpc and
azimuth $50\deg \la \phi \la 130\deg$, we find a significant increase in
the flaring. The strongest deviation from the average flaring is at
$90\deg \la \phi \la 110\deg$. It has already been noted by
\citet{Levine2006a} that this region is special. Inspecting the survey
data, we find a highly disrupted \hi~ distribution with some evidence
that the \hi~ in this range may be significantly perturbed (see
Figs. \ref{Fig_az110} \& \ref{Fig_az110_v220}). We exclude this sector
from the average over the rest of the disk.

Systematic deviations in the flaring between northern and southern part
of the disk are explainable if either the mass distribution is
non-axisymmetric or the state of the ISM is deviant for a part of the
disk. Explaining the asymmetries for $R \ga 25$ kpc as merely due to the
mass distribution leads to a dark matter model that in the southern
part has three times the mass of the northern hemisphere (the fits S and
N1 in Figs. \ref{Fig_flare_south} \& \ref{Fig_flare_north}). Correspondingly, the rotation velocities at $R \ga 25$ kpc should
vary by 50 \kms. This is not observed so we exclude this
possibility. 
 
\begin{figure}[!th]
   \centering
   \includegraphics[angle=-90,width=9cm]{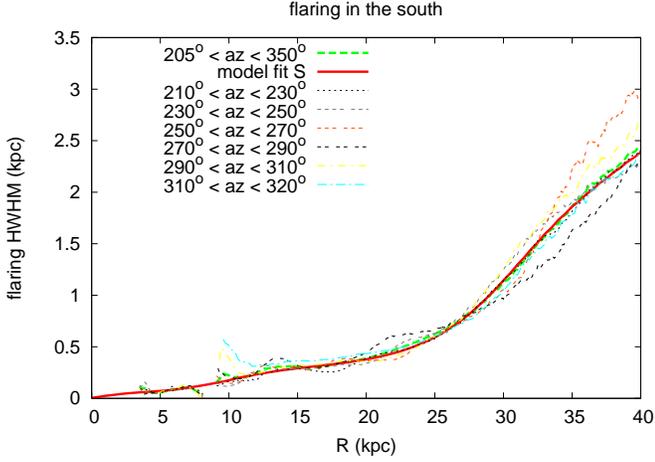}
\caption{Flaring of the southern part of the Galactic disk. The average
flaring for $205\deg < \phi < 350\deg$ (dashed green) is compared with
the model fit (thick red). Flaring curves, derived for individual 20\deg
\ wide sectors, show good internal consistency of the flaring for the
southern part of the disk. }
         \label{Fig_flare_south}
   \end{figure}

\begin{figure}[!ht]
   \centering
   \includegraphics[angle=-90,width=9cm]{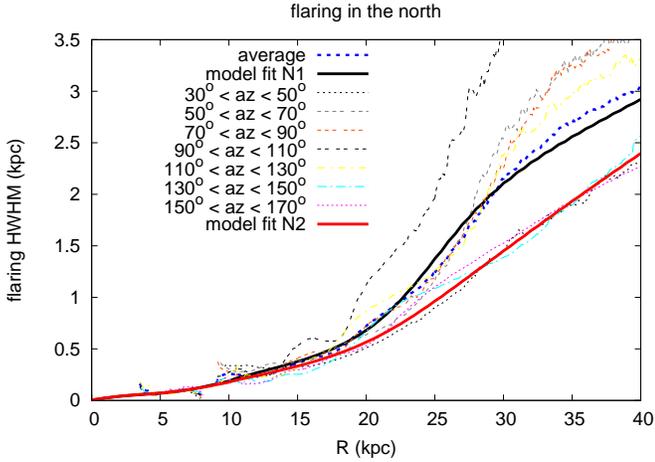}
\caption{Flaring of the northern part of the Galactic disk. The average
flaring, excluding $50\deg < \phi < 130\deg$, is indicated by the dashed
blue line, the corresponding fit (N1) by the black line.  Flaring curves
derived for $20\deg$ wide sectors show systematical deviations for $R >
25$ kpc. The thick red line shows the fit (N2) to the low flaring part
of the \hi~, for $30\deg < \phi < 50\deg$ and $130\deg < \phi <
170\deg$. }
         \label{Fig_flare_north}
   \end{figure}

Taking the opposite point of view that the flaring is affected by
variations in the state of the ISM, we assume for the moment that the
flaring for $50\deg \la \phi \la 130\deg$ is ``atypical'' and may be
disregarded (a more detailed discussion is given in Sect. 9). For the
low flaring part of the northern disk we find that the dark matter disk
has to be somewhat more extended (radial scale length $H_R/2 = 8.75$
kpc) than the disk fitted to the mean flaring used in the previous
sections ($H_R/2 = 7.5$ kpc). For the southern part of the disk
($205\deg < \phi < 350\deg$), we find almost identical parameters, $H_R/2
= 8.5$ kpc.  Fitting the low flaring sectors only (model N2 \& S)
results in a dark matter disk that is, within the errors, axisymmetric.
The total mass of the dark matter disk is affected by the choice of the
model. For best-fit model as displayed in Fig. \ref{Fig_flare_final} we
obtain a mass $ M_{DMD} = 1.8~ 10^{11}$ \msun.  Fitting only the low
flaring part of the \hi~ disk (fits N2 \& S in
Figs. \ref{Fig_flare_south} \& \ref{Fig_flare_north}) leads to $ M_{DMD}
= 2.4~ 10^{11}$ \msun.

Major north/south differences between the models remain and are easily
visible in Figs. \ref{Fig_flare_south} \& \ref{Fig_flare_north} at $R
\sim 25$ kpc.  This is caused by variations in the dark matter ring
radius with best-fit results of $R = 13$ kpc in the north and $R = 18.5$
kpc in the south. Similar variations between 15 and 20 kpc for the
observed stellar ring structure have been noted by
\citet{Ibata2003}. The total mass of the dark matter ring is $ M =
2.2$ to $2.8~ 10^{10}$ \msun. Fits N1 and S imply further that the
southern part of the dark matter ring is three times as massive as the
northern part. Our finding is consistent with known stellar
over-densities in the south \citep[e.g.][]{Bellazzini2006}, suggesting
an association between dark matter ring and stellar distribution. The
ring, however, cannot be associated with the Canis Major over-density.
The radius $R = 18.5$ kpc of the dark matter ring in this direction
implies that this object is located behind Canis Major and instead
associated with ``the One Ring'' discussed previously by
\citet{Newberg2002,Ibata2003,Rocha-Pinto2003} and \citet{Yanny2003}.

North/south differences in the mass distribution, considering the mass
distribution from dark matter ring and disk alltogether (fits N2 \& S),
lead to a lopsided dark matter distribution that is most significant at
$R \sim 25$ kpc. Our fits imply that the rotational velocities at this
distance deviate by $\sim 15$ \kms, in good agreement with the
empirical first-order epicyclic streamline correction derived by
\citet{Levine2006a}. We applied this correction as described in Sect. 2.1. 

\subsection{Spiral structures} 

For a better understanding of the nature of the large-scale fluctuations,
we display in Fig. \ref{Fig_spiral_arms} the relative deviations of the
observed flaring from the axisymmetric model defined as ${\rm
HWHM}_{obs}/{\rm HWHM}_{model}$.}  The north-south asymmetry is obvious, 
but we also find ring-like perturbations and spiral arm
features. Comparing Fig.  \ref{Fig_spiral_arms} with perturbations in
gas thickness as determined by \citet[][Fig. 2]{Levine2006b} shows 
excellent agreement, although very different methods have been
used. This applies to the filtering of the data base for the separation
of disk emission, to the determination of the scale heights, and in
particular to the derivation of perturbations in scale height.
\citet{Levine2006b} use the un-sharp masking method, while we compared
the data with a model. The only agreement between both methods is that
similar rotation curves were applied to converting the observed spectra to
a 3-D volume density distribution. \citet{Levine2006a,Levine2006b} {\it
assume} a flat rotation curve, $v = 220$ \kms, while we {\it derived}
the galactic rotation (Fig. \ref{Fig_rot_curve}) by fitting a mass
model, also ending up with a flat rotation law for $z = 0$ kpc.

First of all we need to discuss the north-south asymmetry of the
flaring. This begins at $ R \ga 15$ kpc, about the same as the distance
at which the warp causes a significant bending. The four-arm spiral
pattern according to \citet{Levine2006b}, indicated in
Fig. \ref{Fig_spiral_arms}, also shows a marked asymmetry with arms wide
open to the north. Structures like these suggest tidal interaction. The
Magellanic system might have caused a dark matter wake, enhancing the
mass distribution towards the south \citep{Weinberg1998}. The tidal
interaction scenario is consistent with the onset of the warp and
higher-mode vertical perturbations at larger distances
\citep{Saha2006,Levine2006a,Weinberg2006}. Large-scale deviations from
axisymmetry, visible in Fig. \ref{Fig_spiral_arms}, are fully consistent
with such a scenario.

\begin{figure}[!th]
   \centering
   \includegraphics[width=9cm]{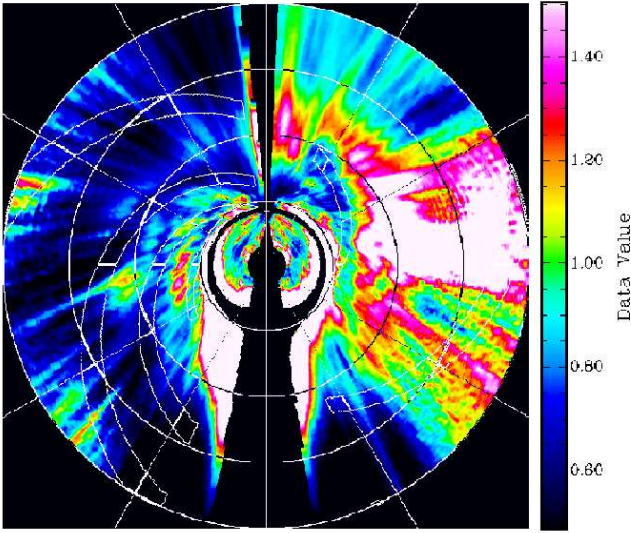}
\caption{Deviations ${\rm HWHM}_{obs}/{\rm HWHM}_{model}$
for the observed flaring relative to the best-fit model. The location of
spiral arms according to \citet{Levine2006b} is indicated. The circles
indicate galactocentric radii of 10, 20, 30, and 40 kpc.}
         \label{Fig_spiral_arms}
   \end{figure}

\begin{figure}[!th]
   \centering
   \includegraphics[angle=-90,width=9cm]{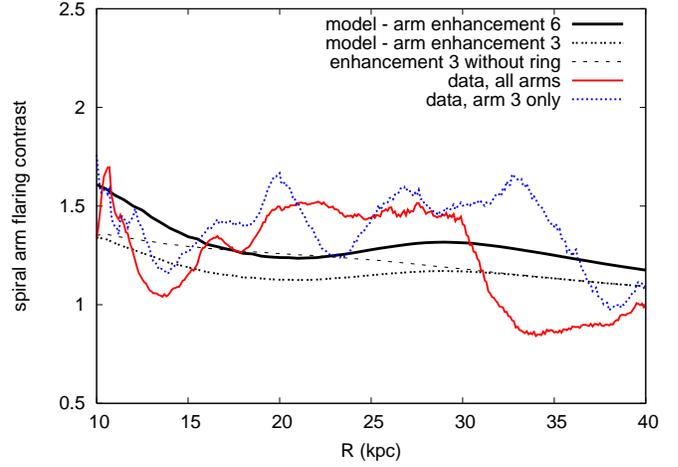}
\caption{Flaring contrast ${\rm HWHM}_{inter-arm}/{\rm HWHM}_{arm}$
between inter-arm and arm regions derived for the arms indicated in
Fig. \ref{Fig_spiral_arms}. The black lines represent model calculations
assuming that the flaring is affected by gravitational forces from
stellar spiral arms. }
         \label{Fig_flare_q}
   \end{figure}

We used the spiral-arm parameters as determined by \citet[][Table
1]{Levine2006b} to verify whether the depressions in scale height along
the spiral arms may be caused by gravitational forces from the stellar
population. For the arm region we set a constant width of $\Delta \phi =
15\deg$ in azimuth, as indicated in Fig. \ref{Fig_spiral_arms}. This is
compared with an inter-arm region with a width of 17\fdg5 on each side
of the arm. Perturbations in direction to the Galactic center or
anti-center, caused by spurious emission, are excluded as indicated.
Figure \ref{Fig_flare_q} plots the ratio ${\rm HWHM}_{inter-arm}/{\rm
HWHM}_{arm}$ determined as an average over all four arms and also for
arm 3 alone. Arm 3 according to \citet[][Fig. 3]{Levine2006b} was chosen
since it is particular well-defined. It runs from $R \sim 12$ kpc, $\phi
\sim 180\deg$ to $R \sim 40$ kpc, $\phi \sim 340\deg$.

We used our model to calculate the flaring perturbations relative to an
axisymmetric model, caused by enhancements of the stellar volume density
within the arms by a factor of 3 or 6 over the axisymmetric
model. A factor of 3 would be expected for well-defined spiral arms, as
in M51 \citep{Rix1993}.  A factor of 6 may be considered as an upper
limit in the case of a four-arm spiral if the stars concentrate in arms
with $\Delta \phi = 15\deg$ in azimuth. 

Comparing the observed contrast with the model, we find that the observed
amplitudes of the stellar spiral arms in the outer parts of the Milky
Way are stronger than expected. This finding, however, is not solid
enough to conclude that the spiral structure in the \hi~ flaring must be
caused by an additional mass component. Gas, according to our model,
contributes only 13\% to the baryonic surface density of the outer
disk. The self gravity of the gas is therefore insignificant. The model
parameters used for the stellar disk, in particular for the exponential
radial scale length of the thick disk \citep[$H_R = 7.5 $ kpc,][]{Chiba},
are highly uncertain. The surface densities for the outer stellar disk
may be underestimated in our model. We conclude that spiral arms
influence the mass distribution out to $R \sim 30$ kpc, possibly even
out to $R \sim 35$ kpc if we consider only arm 3. For comparison,
\citet{Levine2006b} find evidence of a spiral structure out to $R \sim
25$ kpc. The presence of stellar spiral arms appears to be noticeable
out to large distances, approximately $2 R_{25}$.

Figure \ref{Fig_spiral_arms} shows some ring-like fluctuations and
Fig. \ref{Fig_flare_q} gives evidence of radial variations of the
flaring contrast. A close inspection of Figs. \ref{Fig_flare_final},
\ref{Fig_flare_south}, and \ref{Fig_flare_north} also shows some
oscillations if one compares the model flaring with the
observations. Obviously the mean radius of the ring ($R \sim 17.5$ kpc),
as used in our model, is only a crude approximation of a more
complicated pattern of radial substructures in the Milky-Way mass, as
expected for merger events \citep[e.g.][Fig. 11]{Martin2005}.  The tidal
interaction scenario is consistent with the observed radial variations
of spiral arm amplitudes that may arise from interference of a
pre-existing spiral pattern with tidally-induced spiral arms
\citep{Rix1993}. Radial variations in the arm amplitudes in our Galaxy
may be comparable to variations found in M51. An alternative
interpretation of radial fluctuations are ring-like mass concentrations
caused by mergers. Figure \ref{Fig_flare_q} shows that these also influence
the arm/inter-arm contrast.

In summary, we conclude that our model is able to
represent the global properties of the Milky-Way mass distribution. 
Even substructures like spiral arms and radial fluctuations are
explicable. 

\section{Uncertainties} 

In the previous sections we have claimed that a dark matter component
associated with the Galactic disk can fit the observed \hi~ flaring
best, but at least we have found no easy way to match a standard model to
the flaring data.  The best conventional model, a highly prolate
spheroid ($ 2 \la q \la 64$), appears unlikely since this is not what we
expect from $\Lambda$CDM models. Could it be that the analysis is
systematically biased?

As discussed in the previous section, the gravitational field of the
Milky Way must have significant non-axisymmetric contributions.
Accordingly, the Galactic \hi~ disk cannot be in an axisymmetric
equilibrium state. In addition, stellar activities may lead to a
fountain flow and mass may be accreted from outside \citep{HVCbook}. As
part of our analysis we estimate the amount of gas in non-circular
motion for each position in the Galactic plane, thus either representing
out- or inflow. We exclude this gas from our analysis, on average 3\% to
5\% of the total \hi. The remaining uncertainties are most probably below a
percent level, too low to bias derived flaring curves significantly.

Most critical for deriving halo properties is the modeling of the \hi~
scale height. The conventional approach is to assume a single-phase
medium with a velocity dispersion of $\sigma \sim 9.2$ \kms
\citep{OM00,OM01,Narayan2005}. We considered a more sophisticated
two-phase medium with two different characteristic scale heights
\citep{Dickey1990}, defined by the turbulent motions within the \hi~
gas. Using the Leiden/Dwingeloo Survey \citep{Atlas1997}, \citet{KK1998}
determined the average exponential scale height of the CNM to 150 pc and
the scale height of the WNM to 400 pc (see also Table 1 in
\citetalias{Kalberla2003}). The total scale height of the \hi~ gas layer
is in general given by the mix of both phases and therefore depends on
the properties and the composition of the two phases throughout the
disk.

The conditions for the existence of a two-phase medium have recently
been discussed by \citet{Wolfire2003}. These authors conclude that the
CNM and WNM phases must coexist over most of the Galactic disk, at least
out to radii of 16--18 kpc. Basically, such a two-phase medium can
exist as long as the turbulent pressure of the WNM is high enough to
drive parts of the \hi~ gas into a cold dense phase. We repeated the
calculations by \citet{Wolfire2003} of the parameters of the \hi~ gas
distribution. We find, for large radial distances, a 50\% higher
turbulent pressure than used by \citet{Wolfire2003}. This certainly
supports a two-phase medium at large distances. From a Gaussian analysis
of HVCs, \citet{Kalberla2006} find clear evidence of a two-phase medium
in complex H ($R \sim 30$ kpc), the Magellanic stream, and the leading
arm ($R \sim 50$ kpc). In addition they argue for a high pressure to the
hot halo gas phase. We conclude that there is strong evidence that the
\hi~ gas has a multiphase composition throughout most of the Milky Way
disk. Any reservations that the CNM must be absent beyond the extent of
the stellar disk \citep[e.g.][]{OM00,OM01} appear to be outdated.

{ The extended strongly flared region, visible in
Fig. \ref{Fig_spiral_arms} at $ R \ga 25$ kpc and $50\deg \la \phi \la
130\deg$, deviates significantly from the average \hi~ disk gas. This
region has particularly low surface and volume densities and it is
strongly warped.  It is quite possible that the \hi~ gas is out
of equilibrium there and that the CNM phase may be under-abundant. Detailed
investigations would be necessary to verify this, but they are clearly
beyond the scope of this investigation. }

Figure \ref{Fig_flare_HI_inner} displays scale heights of our model
individually for the CNM and WNM phases, but also for the two-phase
mix. Note that the scale height of the two-phase medium is only slightly
larger than the scale height of the CNM. { A WNM gas phase without
any cold \hi~ gas would result in a scale height that is twice as big as
the scale height for the CNM/WNM mixture. This factor of two defines an
upper limit for an excess in flaring caused by the transition of a
multiphase gas into a WNM-dominated phase. Except for the range $90\deg
\la \phi \la 110\deg$, such a factor applies to the region $ R \ga 25$
kpc and $50\deg \la \phi \la 130\deg$, described in more detail in
Sect. 8.1. For the rest of the disk we have no indications that there
are significant changes in the \hi~ composition for $R \ga $ \rsuns so 
use a constant column density ratio for the CNM and WNM phases
throughout the Milky Way. }

Calculating the scale heights from the mass model (Eq. 7 in
\citetalias{Kalberla2003}), we are faced with the problem that
individual gas components are not only supported by their own
turbulence, but are also affected by the pressure of magnetic fields and
cosmic rays. Following \citet{Parker66,Parker69}, the support can be
described as an increase in the velocity dispersion by a factor
$\sqrt{1+\alpha+\beta}$, where $\alpha$ stands for the increase due to
the magnetic field and $\beta$ due to the cosmic-ray component.
A simultaneous determination of the 3-D
distributions of gas, magnetic fields, and cosmic rays is needed to
derive $\alpha$ and $\beta$. From observations, the distribution of
\hi, the diffuse ionized gas \citep{Reynolds1997}, and the $10^{6.15}$ K plasma
\citep{Pietz1998}, as well as the synchrotron radiation \citep{Haslam1982}
and the $\gamma$-ray emission \citep{Fichtel1994}, are available for this
purpose.  Based on a correlation of these datasets, \citet{KK1998}
conclude that, for the CNM and the WNM gas phases, only a limited
support is expected from a turbulent magnetic field component close to
the disk with $\alpha \sim 1/3$. Accordingly, this leads to an effective
increase in the velocity dispersion by a factor of $ \sim 1.15$. For
comparison, \citet{OM00,OM01}, arguing that the pressure support can
be derived from the scale heights of gas, estimate a value of $\sim
1.06$ for the increase in the effective velocity dispersion due to
magnetic fields and cosmic rays.

Our algorithm allows using separate scale heights for the CNM and WNM, but
we find no reason to vary the CNM/WNM ratio or the pressure support from
magnetic fields and cosmic rays with galactocentric radius. Our solution
is close to a single-component model with an effective constant velocity
dispersion of 8.3 \kms, somewhat low in comparison to the preferred
value of 9.2 \kmss of \citet{OM00,OM01}. These authors explored the
parameter space, affecting \hi~ flaring for various dispersions at a
single constant galactocentric radius $R =2$ \rsun. In contrast, we use
a constant dispersion for a broad range of distances, and the question
arises whether this is justified. It is very difficult to estimate how
far the isothermal assumption may lead to possible biases. However, we
may turn the question around and ask how far we would need to modify our
ISM model assumptions if we want to save the paradigm of the
Milky Way surrounded by a spheroidal halo.

For a spheroidal mass model without a ring and $q = 1$
(Fig. \ref{Fig_flare_noDMD_noring}) we would need to increase the
estimate for the flaring at $R \sim 17$ kpc by 40\% and, at the same
time, to decrease the estimate at $R \sim 35$ kpc by 40\%. Pressure
support from magnetic fields might be missing at large distances, which
would bring the scale down by 15\%. In turn, we need to assume that
velocity dispersion of the \hi~ gas is overestimated by 25\%. Given the
uncertainties, such a solution cannot be excluded. The arguments are
the opposite at $R \sim 17$ kpc. Here we need to boost the turbulent gas
pressure by 40\%. However, the question immediately arises as to what kinds of
processes would be available to feed this additional turbulence at
distances $12 \la R \la 22$ kpc. This region is largely outside the
stellar disk, hence we expect no support from stellar activities.
Figure \ref{Fig_flare_noDMD_noring} shows that highly prolate spheroids
fit best at large distances. However, the problem remains to explain a
turbulent pressure support at $ R \la 25$ kpc. Alternatively, highly
oblate models fit the observations well up to $ R \la 25$ kpc, but it
is then a serious problem to explain the observed flaring in the
outskirts of the Milky Way. 

Other mass models without dark matter in the Galactic disk might be
considered. \citet{Narayan2005} advocate a model according to
Eq. \ref{Eq_sphere} with a power index $p = 2$. Such a model fits the
\hi~ flaring well. The total halo mass is $ 2.8 \ 10^{11} $ \msun,
however, much too low to be acceptable. This model has most of its mass
within $R \la 50$ kpc. Adding an extended halo to the
\citet{Narayan2005} model would result in a bimodal dark matter model,
very similar to our case.


\section{Summary} 

The aim of our contribution is to derive an \hi~ volume density distribution
$n(R,Z,\phi)$ for the Milky Way. The main driver behind this is the
availability of the LAB survey, a new all-sky \hi~ survey
\citep{Kalberla2005}. In particular, we expect that the low internal
errors of this survey may lead to higher fidelity in the derived
distribution. 

Previously, this survey was used by \citet{Levine2006a,Levine2006b}, who
{\it assume} a constant $v = 220$ \kmss rotation law. In performing the
$T_B(l,b,v)$ to $n(R,z,\phi)$ conversion we notice, however, that
uncertainties in the Milky Way rotation curve are large enough to make
the results ambiguous. We have tried to solve these problems by matching
mass models of the Milky Way to the observations. The strategy is to
obtain a {\it self-consistent} solution. We generated mass models that
satisfy both the Poisson and the Boltzmann equation at the same time.
To apply a model we require that predictions from the mass model need to
agree with the derived volume density distribution. In particular, we
used the global flaring of the \hi~ gas layer to decide on the
consistency of a model. The link between derived volume density
distribution and flaring according to a mass model is the rotation
curve, requiring a {\it self-consistent} mass distribution. Thus, our
procedure is based on a closed loop.

\subsection{Properties of the mass model}

{ The Galaxy is embedded in an isothermal dark matter halo with a
core radius of 35 kpc and a total mass of $ M = 1.8~ 10^{12}$ \msuns
within 350 kpc. To fit the global flaring of the \hi~ disk, we need dark
matter associated with the disk in the form of a self-gravitating
exponential disk with a radial scale length of $7.5-8.75$ kpc and a
mass $ M_{DMD} = 1.8$ to $2.4~ 10^{11}$ \msun. To explain \hi~ flaring
at $13 \la R \la 20$ kpc, we need an additional component. A dark matter
ring at a radius $13 \la R \la 18.5 $ kpc contributes $ M = 2.2 $ to $ 
2.8~ 10^{10}$ \msun. About 2/3 of the ring mass is located in the south at
$R \sim 18$ kpc, while the remaining part of the ring mass is in the
north at $R\sim 13$kpc.

According to our best-fit model (the average flaring as displayed in
Fig. \ref{Fig_flare_final}), the total mass of the Milky Way disk within
50 kpc is $ M = (2.9\pm.1)~ 10^{11}$ \msun. However, the state of the
ISM at $R \ga 22$ kpc is rather uncertain (Sects. 8 \& 9). The northern
part of the outer disk may be dominated by the WNM phase. Taking this
into account by fitting the low flaring part of the \hi~ disk only (fits
N2 \& S in Figs. \ref{Fig_flare_south} \& \ref{Fig_flare_north}) leads
to $ M \sim 3.5~ 10^{11}$ \msun, probably an upper limit.   }

\subsection{Global properties of the \hi~ disk}

Our best-fit model results in a rotation curve that is essentially flat
for $R \la 27 $ kpc and falling off gradually for larger distances. Our
derived \hi~ volume density distribution resembles previous results
obtained for a constant rotation \citep{Voskes1999,Levine2006a}, but we
derived cleaner pictures of the Milky Way \hi~ gas distribution and
extended the investigations to larger distances.  A significant
improvement is obtained by dropping the assumption that the gas is
barotropic, hence in cylindrical rotation.

We derived a well-defined exponential \hi~ disk with $ n(r) \sim {\rm
exp}(-R/3)$. The \hi~ surface density distribution is also exponential
and has a radial scale length of 3.75 kpc.  The \hi~ disk flares
strongly. For $R \sim 4$ kpc the HWHM scale height of the \hi~ layer is
0.06 kpc, increasing to 2.7 kpc at $ \sim 40$ kpc. 

The \hi~ disk shows systematic north-south asymmetries for $R \ga 15$
kpc, apparently correlated with the warp. The warp has well-defined
Fourier modes 0 to 2 with slowly variable phases. The mean flaring in
the northern part is up to a factor of 2 larger than the flaring in the
south.

\section{Discussion}

The dark matter disk, proposed by us, appears to conflict with
results from HIPPARCOS \citep{HF2000,HF2004,Korchagin2003}. The
HIPPARCOS data are believed to imply that there is no evidence
for significant amounts of dark matter in the disk. However, the flaw
in this frequently used statement is that a definition of the
disk is missing.  Constraints from the stellar surface
densities define only a local lower limit of $2 - 3$ kpc for the scale
height of a dark matter component associated with the disk.  The dark
matter disk in our model has a local scale height of $h_z = 4$ kpc and
matches all HIPPARCOS constraints. Models without any dark matter within
the disk are not consistent with the observed gaseous flaring, in
particular for distances $ R \ga 22$ kpc.

A dark matter disk, or alternatively a highly prolate spheroid, explains
the global flaring of the Galactic \hi~ disk. But there remains a marked
depression in the derived flaring at distances $15 \la R \la 20$ kpc. To
explain this it is in any case necessary to add a massive ring.

\subsection{Lord of the rings}

A longstanding discrepancy, discussed in Sect. 1, is that the observed
rotational velocities for $R \ga $ \rsuns conflict with mass
models. To solve this problem, \citet{Binney1997} propose that most of
the tracers that appear to be at $R \ga 11 $ kpc might actually be
concentrated within a ring at $R \sim 14$ kpc. Stellar over-densities
suggesting a ring-like concentration were found a few years later
by several groups independently. Perhaps the most concise description
of this feature was given by \citet{Ibata2003}; there is a giant
stellar ring at $ 15 \la R \la 20 $ kpc with a radial thickness of 2
kpc and a vertical scale height of $\sim 0.75$ kpc. The total stellar
mass is estimated to be between $ 2 \ 10^{8}$ \msuns and $10^9$ \msun. 

Such a mass, however, is insufficient for explaining the observed
flaring. Independent evidence of a massive ring was given recently by
\citet{deBoer2005}. EGRET observations show excess in
the diffuse Galactic $\gamma$-ray background. A possible explanation is
that the giant stellar ring is associated with a dark matter ring.  We
included this mass component in our dark matter disk model but noticed
that the mass of such a ring, initially proposed by \citet{deBoer2005},
needs to be reduced, and the position and extension also needs to be
slightly modified to explain the \hi~ flaring. A fully satisfying
self-consistent solution of our mass model and the average derived
density distribution is obtained for a dark matter ring at $ R \sim 17.5$
kpc with a radial extension of 5 kpc and a mass $ M = 2.3~ 10^{10}$
\msuns (see Fig. \ref{Fig_flare_final}). The associated rotation curve
is almost flat out to $R \la 27$ kpc. This extended dark matter
ring, associated with a somewhat less extended stellar ring and some
enhancements in \hi~ is completely consistent with the proposal by
\citet{Binney1997} that distance errors may mimic a rising rotation
curve for $R \ga 11$ kpc.

The \citet{deBoer2005} paper explains the EGRET excess emission as due to dark
matter annihilation (DMA). From the energy spectrum of the excess
radiation, they deduce a WIMP mass between 50 and 100 GeV. The
significance of the EGRET excess emission was confirmed by
\citet{Bergstroem2006}, but these authors question that the excess can
be caused by DMA. They model the decay processes proposed by
\citet{deBoer2005} and derive that such a DMA would be associated by a
significant antiproton flux.  The expected flux is well above
observational limitations, so they conclude that the signal cannot be
caused by DMA. \citet{Bergstroem2006} propose to explain the EGRET
excess by enhancements in the baryonic matter distribution within the
area covered by the ring.

As a crude order-of-magnitude estimate, the column density of the ring
along the line of sight is comparable to the total column density of
the observable \hi~ disk in the same direction. For such baryonic
matter contained within the usual gas phases, a {\it huge} EGRET
signal would be expected. This is clearly incompatible with the
background signal observed by \citet{deBoer2005}, where any possible
baryonic matter within the ring must be largely hidden from the
observer by self-shielding effects.

Baryonic dark matter in the form of cold, dense molecular gas clumps
(clumpuscules), associated with the \hi~ disk, was proposed in
\citet{Pfenniger1994a} and \citet{Pfenniger1994b}. Such clumps are hardly
detectable optically or with radio telescopes. They are self-shielding
against $\gamma$-ray emission if their diameters are sufficiently
small, e.g. $\la 10$ AU for a clump mass of $10^{-3}$ \msuns
\citep{Kalberla1999,Ohishi2004}. Excess $\gamma$-ray emission,
correlated with local gas and dust was found by \citet{Grenier2005}.
These investigations suggest that some ``dark gas'' is associated with
\hi~ gas, distributed like an envelope around CO-traced molecular
gas. The total mass of this ``dark gas'' is comparable to the mass of
the molecular gas, negligible in comparison to the dark matter ring.

\subsection{Origin of the dark matter ring and disk}

Our mass model is axisymmetric. However, the derived flaring shows
significant asymmetries (Figs. \ref{Fig_flare_south} to
\ref{Fig_spiral_arms}). Spiral structure exists out to $ R \sim 30$ kpc,
and we find strong radial variations and a pronounced north-south
asymmetry. Typical scales for systematic fluctuations appear to be in
the range 3 to 15 kpc.

The Milky Way is not a unique galaxy. Rotation curves of other 
galaxies show similar wiggles; perhaps the most impressive collection 
is given by \citet{Sofue1999}. It appears hard to explain
all these fluctuations as caused by an extended dark matter halo. The
current understanding is that the halos represent a rather smooth
distribution. Despite density enhancements at the center (cusps), no
substructures on kpc scale are expected. Clumpyness appears instead to be
associated with dissipative matter \citep{Combes2002}.

The dark matter ring may provide a key to understanding these
fluctuations. { The ring, located at $13 \la R \la 18.5 $ kpc, has a
FWHM extension of 12 kpc.  Associated stellar streams appear to be more
sharply bounded in distance, in FWHM typically 2 kpc to 4 kpc
\citep{Rocha-Pinto2003,Martinez2005}. \citet{Newberg2002} determine an
upper limit of 6 kpc.} Simulations of the accretion of a dwarf onto the
Galaxy show that the most prominent tidal arms are also spatially
well-defined \citep{Martin2005}.

If the giant stellar ring is caused by the disruption of a dwarf galaxy,
it appears reasonable to also associate the dark matter ring with the
dwarf. The mass-over-light ratio for the progenitor galaxy would be in
the range 24 to 120, with the dark matter distribution more extended by
a factor of 4. This case resembles the Sagittarius dwarf, as modeled by
\citet{Ibata1997} and \citet{Ibata1998}. These authors needed a similar
dark matter halo to stabilize the dwarf against the Galactic tides.

\citet{Helmi1999} argue that the stellar halo may have been built up by
disrupted satellites. Their calculations suggest that the stellar halo
could consist of 300 to 500 streams. Assuming that these streams are
associated with dark matter would imply a more or less smooth spatial
distribution of dark matter debris within the Galaxy. The hypothesis
that satellites may have considerable amounts of dark matter is
supported by \citet{Gilmore2006}, who argue that Milky Way dwarf
spheroidals are dark matter dominated by halos extending possibly to kpc
scales with typical mass-to-light ratios of a few
hundreds. \citet{Hayashi2003} essentially come to the same conclusion
when simulating the structural evolution of substructures.  From
dissipation-less simulations on structure formation, a roundish halo is
expected \citep{Allgood2006}. The dark matter disk advocated by us
implies a strongly anisotropic distribution for the progenitors of the
stellar streams. The orbits for the majority of the accreted dwarfs must
have been highly correlated with the plane of the Milky Way disk.

\citet{Willman2004} discuss the completeness of the census of Milky
Way companions with respect to Galactic latitudes. For a random
distribution, they estimate an incompleteness as large as 33\%.
Up to 10 satellites at low latitudes, possibly causing 300 to 500
streams, if captured, could be waiting for detection or are already
captured by the Galaxy. The recent detection of five new Milky Way
companions \citep{Belokurov2007} implies that this is only a lower limit.
\citet{Abadi2003} demonstrate for disk
galaxies that the bulk ($\sim 60$\%) of the thick disk consists of the
tidal debris from satellites whose orbital plane was coincident with the
disk and whose orbits were circularized by dynamical friction prior to
full disruption. If these satellites contained also dark matter, one
would expect that this material would have built up a thick disk-like
dark matter component. The shape of the resulting dark matter disk would
depend on the accretion history and on the nature of the dark matter.


\begin{acknowledgements} 
  P. Kalberla and L. Dedes acknowledge financial support from the
  Deutsche Forschungsgemeinschaft, grant KA1265/5-1, U. Haud from the
  Estonian Science Foundation grant no. 6106. A previous version of this
  paper was submitted to ApJ, but rejected. We have taken the comments
  of the anonymous referee then into account and thank C. Heiles for
  encouraging comments in the early stages. The project got re-animated
  after getting knowledge of the dark matter ring hypothesis 
  \citep{deBoer2005}, for which we thank G. J\'ozsa and F. Kenn.
  We thank L. Blitz and the A\&A referee for constructive criticism,
  K.S. de Boer and C. Vlahakis for a critical reading of the manuscript,
  J. Brand, S. Malhotra, and J. Wouterloot for making
  observational results available.
\end{acknowledgements}

\end{document}